\documentclass[12pt,amsmath,amssymb]{article}
\usepackage{amssymb}
\usepackage{dsfont}
\usepackage{bm}
\usepackage{mathrsfs}
\usepackage{amsmath}
\usepackage{graphicx}
\numberwithin{equation}{section}

\begin{document}

\title{\sc Quantum Physics of Simple Optical Instruments}
\author{Ulf Leonhardt\\
School of Physics and Astronomy, University of St Andrews,\\
North Haugh, St Andrews, Fife, KY16 9SS, Scotland}
\maketitle
\begin{abstract}
Simple optical instruments are linear optical networks
where the incident light modes are turned into equal
numbers of outgoing modes by linear transformations.
For example, such instruments are beam splitters, multiports,
interferometers, fibre couplers, polarizers, gravitational lenses,
parametric amplifiers, phase-conjugating mirrors
and also black holes.
The article develops the quantum theory of
simple optical instruments and applies the theory to
a few characteristic situations, to the
splitting and interference of photons and to the
manifestation of Einstein-Podolsky-Rosen correlations
in parametric downconversion.
How to model irreversible devices
such as absorbers and amplifiers is also shown.
Finally, the article develops the theory of Hawking radiation
for a simple optical black hole.
The paper is intended as a primer,
as a nearly self-consistent tutorial.
The reader should be familiar with
basic quantum mechanics and statistics, and perhaps with optics
and some elementary field theory.
The quantum theory of light in dielectrics serves as
the starting point and, in the concluding section,
as a guide to understand quantum black holes.

\end{abstract}
\newpage

\section{Introduction}

Consider a semi-transparent mirror, the glass of your window, for example.
The mirror partially reflects light and is partially transparent.
If the material of the mirror is not absorptive the incident light is
exactly split into the reflected and the transmitted component.
Now, light consists of photons, of indivisible light particles.
How does this beam splitter act on individual photons \cite{Brendel,Paul}?
How are photons split?
Or, in another experiment \cite{Hong},
suppose you take a semi-silvered mirror with 50:50
transmission-reflection ratio.
You let exactly one photon propagate towards the frontside of the mirror
and you send another single photon towards the backside,
and let them interfere.
The interference between two light beams depends on their relative phase.
If the phase difference is right, the sum of the two incident beams,
the two photons, emerge behind the mirror
and if they interfere with the opposite phase they appear in front of the mirror.
But single photons are not supposed to carry a precise phase,
because phase is a wave property and individual photons are particles.
So what happens \cite{Hong,Paul}?

Such conundrums are beginning to occupy people's minds for other reasons
than purely academic curiosity, because they may fundamentally alter
our approach to secure data communication \cite{Bruss,Gisin}.
Suppose that Alice wants to send a secret message to Bob,
carried by photons in a glass fiber \cite{Tittel} or through space \cite{Kurtsiefer}.
Eve, the eavesdropper, tries to intercept the message without getting caught.
Clearly, in order to do so, she must probe the stream of messenger photons,
for example using something like a beam splitter.
However, knowing the principal quantum effects of beam splitters,
Alice and Bob may infer from the statistical properties of the transmitted photons
that their secrecy is at risk and may discard the communication channel.
In a more sophisticated eavesdropping attempt,
Eve might amplify the incident light and extract,
by beam splitting, the bits that are sufficient for her.
The rest, with the same amplitude as the original, is transmitted to Bob.
Would Alice and Bob notice Eve's subtle interception?

The quantum physics of simple optical instruments, such as beam splitters
and amplifiers, is clearly important when the quantum nature of light
is used for practical (or academic) purposes.
Equally importantly, some seemingly simple questions about light and
the relatively simple experiments to demonstrate their intriguing answers
do both illuminate and challenge our understanding of the quantum world
\cite{Paul}.
Furthermore, in studying the quantum physics of simple optical instruments
we may see connections to a much wider and sometimes quite exotic
range of physics.
For example, the Hawking radiation of black holes \cite{Hawking1}
is related to the quantum optics of moving media and
combines aspects of beam splitters and amplifiers.

In this article we analyze the principles of quantum-optical
networks where two or more beams of light interact with each
other. The networks are assumed to be linear in the sense that
the output amplitudes depend on the input amplitudes by a linear
transformation, although the physics of such networks may be
based on non-linear optics \cite{Bloembergen,Perina,Shen}.
The article is
intended as a primer, as a nearly self-consistent tutorial, rather
than a literature survey. The reader should be familiar with
basic quantum mechanics and statistics, and perhaps with optics
and some elementary field theory. However, when appropriate, we
quote the major mathematical results needed, instead of deriving
them, for not overburdening this article. We use the notation of
the book \cite{Leonhardt} and some of the basic results of quantum
optics explained there. First we begin with an example, quantum
light in planar dielectrics. Then we extend the central features
of this example to quantum-optical networks in general. We
describe the quantum physics of such networks in the Heisenberg
picture and in the Schr\"odinger picture, and with the help of
quasiprobability distributions such as the Wigner function
\cite{Leonhardt}. In Sec.\ 4 we develop the quantum
optics of the beam splitter, because this simple device is the
archetype of all passive optical instruments, and because the
beam splitter is capable of demonstrating many interesting
aspects of the wave-particle dualism. In Sec.\ 5 we analyze
absorbers and amplifiers, which are irreversible devices, and show
how they can be described by effective models, such as fictitious beam
splitters and parametric amplifiers. In Sec.\ 6 we develop
the quantum theory of parametric amplifiers and
phase-conjugating mirrors. Parametric amplifiers have been widely
used to experimentally demonstrate the nonlocality of quantum
mechanics in versions of the Einstein-Podolsky-Rosen paradox
\cite{Bell,EPR}. Section 7 returns to the starting point, to
quantum light in dielectric media. We show how moving media may
establish analogs of black holes and we analyze the
essentials of Hawking radiation \cite{Hawking1}. Throughout
this article, whenever possible, we try to use models that are
simple but not too simple.

\section{Quantum optics in dielectrics}

Many passive optical instruments such as lenses or beam splitters
consist of dielectric materials like glass that influence the
propagation of light without causing much absorption.
Dielectrics are linear-response media --- their effect on light
is proportional to the electromagnetic field strengths.
The field induces microscopic dipoles in the atoms constituting
the dielectric medium.
The dipoles constitute macroscopic electric polarizations and
magnetizations that are proportional to the applied
electromagnetic field and which act back onto the field.
An isotropic medium is characterized by two spatially dependent
proportionality factors, the electric permittivity $\varepsilon$
and the magnetic permeability $\mu$ \cite{Jackson,LL8}.
We assume that the medium is not dispersive and not dissipative.
In this case both $\varepsilon$ and $\mu$ are real and do not depend
on the frequency of light within the frequency window we are
considering. The square root of the product of $\varepsilon$ and $\mu$
gives the refractive index that describes the degree to which
the phase velocity of light in the medium deviates from the speed
of light in vacuum, $c$.

The quantum theory of light in dielectrics has been subject
to a substantial literature summarized to some extent in Refs.\
\cite{Barnett,GlauberLewen,Knoell1,Knoell2,VWW}.
Traditional quantum optics is the subject of the recent books
\cite{Bachor,BR,Cohen,Leonhardt,Loudon,MandelWolf,Perina,PerinaEd,Schleich,SZ,VWW,WM}.
Here we consider the simplest possible case where the dielectric
functions $\varepsilon$ and $\mu$ vary in one direction of space only,
say in $x$ direction. Furthermore, we assume that the electromagnetic
waves propagate in this direction as well and we select one of the two
polarizations of light.
In this way we arrive at an effectively one-dimensional model.

\subsection{Classical fields}

Consider the classical electromagnetic field characterized by the
electric field strength $E$ and by the magnetic $B$ field in SI units.
The electromagnetic field obeys the Principle of Least Action
\cite{LL2} with the Lagrangian density \cite{GlauberLewen}
\begin{equation}
\label{eq:lagrange}
{\mathscr L}=\frac{\varepsilon_0}{2}
\left(\varepsilon E^2 -\frac{c^2}{\mu} B^2\right) \,.
\end{equation}
In one spatial dimension the vector potential in Coulomb gauge
\cite{Cohen} is effectively a scalar field $A$ with
\begin{equation}
\label{eq:vectorpotential}
E=-\partial_t A\,,\quad B=\partial_x A \,.
\end{equation}
Throughout this article we abbreviate partial-differentiation operators
such as $\partial/\partial t$ and $\partial/\partial x$
by $\partial_t$ and $\partial_x$.
We obtain from the Lagrangian (\ref{eq:lagrange})
the Euler-Lagrange equation
\begin{equation}
\label{eq:wave}
\frac{1}{\varepsilon}\,\partial_x \frac{1}{\mu}\,\partial_x A -
\frac{1}{c^2}\, \partial_t^2 A = 0\,,
\end{equation}
the wave equation of light in one-dimensional media at rest.
We define the scalar product between two fields with
vector potentials $A_1$ and $A_2$ as
\begin{equation}
\label{eq:scalar}
\left(A_1,A_2\right) \equiv \frac{i\varepsilon_0}{\hbar}
\int \left(A_1^*\,\partial_t\, A_2 - A_2 \,\partial_t \,A_1^*\right) \varepsilon\,
\mathrm{d}x\,.
\end{equation}
As usual, $\hbar$ denotes Planck's constant divided by $2\pi$.
The scalar product (\ref{eq:scalar}) is a conserved quantity
as a consequence of the wave equation  (\ref{eq:wave}),
\begin{equation}
\partial_t\left(A_1,A_2\right) =0\,,
\end{equation}
and the product plays an important role in the mode
decomposition of quantum light.

\subsection{Quantum fields}

According to the quantum theory of light
\cite{Cohen,Loudon,MandelWolf}
the vector potential is regarded as a quantum observable,
as a Hermitian operator $\hat{A}$ that depends on space and time.
We represent $\hat{A}$ as a superposition of modes
\begin{equation}
\label{eq:modeex}
\hat{A}(x,t) = \sum_k \Big(A_k(x,t)\,\hat{a}_k+
A_k^*(x,t)\,\hat{a}^\dagger_k \Big)\,.
\end{equation}
The mode functions $A_k$ satisfy the classical wave equation
(\ref{eq:wave}). They describe how single light quanta, photons,
propagate in space and time, given the initial and boundary conditions
that determine the particular $A_k$.
They also describe how coherent states
\cite{Leonhardt,Loudon,MandelWolf}
propagate, states that describe classical light fields.
The mode functions characterize the classical, wave-like,
properties of light, whereas the quantum amplitudes $\hat{a}_k$
describe the quantum features of light.
We require that the mode functions are orthonormal with respect
to the scalar product (\ref{eq:scalar}),
\begin{equation}
\label{eq:orthonorm}
\left(A_k, A_{k'}\right) = \delta_{kk'}\,,\quad
\left(A_k^*, A_{k'}\right) =0\,.
\end{equation}
As a consequence, the commutator relation between $\hat{A}$
and the canonically conjugate momentum \cite{Weinberg} implies
\begin{equation}
\label{eq:comm}
[\hat{a}_k, \hat{a}_{k'}^\dagger] =  \delta_{kk'}
\,,\quad
[\hat{a}_k, \hat{a}_{k'}] =  0
\,.
\end{equation}
Therefore, light quanta are bosons \cite{Weinberg},
{\it i.e.}\  quanta of harmonic oscillators,
with annihilation operators $\hat{a}_k$ and creation operators
$\hat{a}_k^\dagger$.
Each mode of light represents an electromagnetic oscillator,
the best harmonic oscillators currently known
(with the smallest anharmonicity,
generated by the vacuum polarization
due to electron-positron pairs \cite{HeisenbergEuler,Milonni},
an effect beyond our model).
Furthermore, if we choose monochromatic mode functions with
frequencies $\omega_k$, we obtain for the field energy
\begin{equation}
\int \frac{\varepsilon_0}{2}
\left(\varepsilon \hat{E}^2 +\frac{c^2}{\mu} \hat{B}^2\right) {\mathrm d}x =
\sum_k \hbar\omega_k \left(\hat{a}_k^\dagger\hat{a}_k + \frac{1}{2}\right)\,.
\end{equation}
Each mode contributes to the total energy
as the energy quantum $\hbar\omega_k$ times the photon number
$\hat{a}_k^\dagger\hat{a}_k$.
The additional vacuum energy $\sum_k\hbar\omega_k/2$
does not depend on the quantum state of the electromagnetic field,
but it may depend on the boundary conditions,
giving rise to the Casimir force \cite{Barash,Chan,Lamoreaux,Milonni}.

In quantum optics, the mode operators $\hat{a}_k$ are frequently represented
in terms of the quadrature operators $\hat{q}_k$ and $\hat{p}_k$
\cite{Leonhardt}
\begin{equation}
\label{eq:quad} \hat{a}_k =
\frac{1}{\sqrt{2}}\,(\hat{q}_k+i\hat{p}_k)\,,\quad \hat{q}_k =
\frac{1}{\sqrt{2}}\,(\hat{a}_k^\dagger+\hat{a}_k)\,,\quad
\hat{p}_k = \frac{i}{\sqrt{2}}\,(\hat{a}_k^\dagger-\hat{a}_k)\,.
\end{equation}
The quadratures play the role of the real and the imaginary parts of the
mode amplitudes. They satisfy the Heisenberg commutation relation
(with $\hbar =1$)
\begin{equation}
\label{eq:heisen} [\hat{q}_k,\hat{p}_{k'}] = i \delta_{kk'}\,.
\end{equation}
The $q$ quadrature appears as the position and the $p$ quadrature as
the momentum of the electromagnetic oscillator represented in a single
mode of light. This correspondence between light amplitudes
and canonically conjugate quantities has found interesting applications
in simultaneous measurements of position and momentum
\cite{Leonhardt,WalkerCarroll,Walker}
and in quantum-state tomography
\cite{Breitenbach,LeoPaul,Leonhardt,Lvovsky,Smithey,Welsch},
because the quadratures can be measured with high precision
in balanced homodyne detection \cite{Abbas,YuenChan},
for further details see Ref.\ \cite{Leonhardt}.

\subsection{Transfer matrix}

Optical instruments act primarily on the classical wave-like properties of light.
In the regime of far-field optics, the instrument is spatially well separated
from the light sources and from the places where the light is detected
or otherwise applied to.
In this situation, we can decompose both the incident light and the outgoing
light into plane waves. The reflection and transmission coefficients
of the incident plane waves characterize the performance of the instrument.
The coefficients constitute the transfer matrix of the dielectric structure.

Consider monochromatic light of frequency $\omega$ propagating
in a one-dimensional lossless dielectric. In a region where the dielectric
functions $\varepsilon$ and $\mu$ do not vary, the mode function $A_k$
is a superposition of waves traveling to the right, $\exp(ikx-i\omega t)$,
and waves traveling to the left, $\exp(-ikx-i\omega t)$, where $k$ denotes
the wavenumber
\begin{equation}
k=\frac{\omega}{c}\,\sqrt{\varepsilon\mu} \,.
\end{equation}
Consider \cite{Bandelow}
\begin{equation}
A_{\pm k} = \frac{1}{2}\,\exp\left({\mp i\varphi}\right) \left(A_k \pm
\frac{1}{ik}\,\partial_x A_k\right)\,,\quad
\varphi = \int k\,\mathrm{d}x\,.
\end{equation}
In the region where the dielectric is uniform, $A_{+k}$ picks out
the coefficient of the right-moving component of $A$,
whereas $A_{-k}$ gives the left-moving part.
When the dielectric functions vary, the $A_{\pm k}$ serve to
identify how the wave coefficients are transferred across the dielectric
structure from a region of asymptotically constant
$\varepsilon_L$ and $\mu_L$ on the left to a region of
(possibly different) $\varepsilon_R$ and $\mu_R$ on the right.
We obtain from the wave equation (\ref{eq:wave})
\begin{equation}
\label{eq:apm}
\partial_x
\left(
    \begin{array}{c}
      A_{-k} \\
      A_{+k}
    \end{array}
\right)
=
\frac{(\partial_x Z)}{2Z}
\left(
    \begin{array}{cc}
      1 & -\exp(+2i\varphi) \\
      -\exp(-2i\varphi) & 1
    \end{array}
\right)
\left(
    \begin{array}{c}
      A_{-k} \\
      A_{+k}
    \end{array}
\right)\,,
\end{equation}
where $Z$ denotes the impedance \cite{Jackson}
\begin{equation}
Z = \sqrt{\frac{\mu}{\varepsilon}}\,.
\end{equation}
In order to get reflection the impedance must vary, as we see from
Eq.\ (\ref{eq:apm}). In the case of perfect impedance matching
$Z$ remains constant, and the structure is guaranteed to be reflectionless,
a result well known from the physics of transmission lines \cite{Jackson}.
To get strong reflection, with, in the extreme case, total reflection caused
by photonic bandgaps \cite{Joannopoulos},
the dielectric structure should periodically vary at
about twice the wave length of light, as we infer from the
oscillating terms in Eq. (\ref{eq:apm}).
We express the general solution of the differential equation (\ref{eq:apm}) as
\begin{equation}
\label{eq:connex}
\left(
    \begin{array}{c}
      A_{-k}(x_2,t) \\
      A_{+k}(x_2,t)
    \end{array}
\right)
=
T(x_2,x_1)
\left(
    \begin{array}{c}
      A_{-k}(x_1,t) \\
      A_{+k}(x_1,t)
    \end{array}
\right)\,,
\end{equation}
where $T(x_2,x_1)$ denotes the transfer matrix from $x_1$ to
$x_2$. The columns of the matrix $T(x,x_1)$ are required to
satisfy the differential equation  (\ref{eq:apm}) with the
initial condition $T(x_1,x_1)=\mathds{1}$. The transfer matrix has
the structure
\begin{equation}
\label{eq:trans}
T =
\left(
    \begin{array}{cc}
      a & b^* \\
      b & a^*
    \end{array}
\right) \,,
\end{equation}
because, if $(a,b)^T$ solves Eq.\ (\ref{eq:apm}) so does $(b^*,a^*)^T$.
Furthermore, the spatial derivative of the determinant of $T$ equals the
spatial derivative of the impedance $Z$. Consequently,
\begin{equation}
\label{eq:det}
|a|^2-|b|^2 = \frac{Z(x_1)}{Z(x_2)}\,,
\end{equation}
and hence
\begin{equation}
\label{eq:invt}
T^{-1} =
 \frac{Z(x_2)}{Z(x_1)}
\left(
    \begin{array}{cc}
      a^* & -b^* \\
      -b & a
    \end{array}
\right) \,.
\end{equation}
The transfer matrix $T(+\infty,-\infty)$ characterizes the far-field
performance of the optical instrument made of the dielectric structure.
On the other hand, the transfer matrix does not directly describe
how the two incident light modes interfere with each other to produce
the outgoing modes.

\subsection{Scattering matrix}

In one spatial dimension, the directions of light propagation are fairly
restricted
--- the incident light can come from the left or from the right of the
dielectric structure.
Waves incident from the left, $A_1^\mathrm{in}$, are partially reflected
and partially transmitted, but beyond the structure the waves must
propagate to the right.
(We drop the mode index $k$ for simplicity.)
Similarly, waves coming from the right, $A_2^\mathrm{in}$,
are purely outgoing towards the left.
We assume monochromatic modes
\begin{equation}
A_1^\mathrm{in} = u_1(x)\,e^{-i\omega t} \,,\quad
A_2^\mathrm{in} = u_2(x)\,e^{-i\omega t} \,,
\end{equation}
and utilize the inverse transfer matrix (\ref{eq:invt}) to define $u_1$
as the spatial component of a wave with the asymptotics
\begin{equation}
\label{eq:as1}
u_1(x) \sim {\cal A}_1
\left\{
\begin{array}{r@{\quad:\quad}l}
\frac{Z_R}{Z_L}\left(a\,e^{ik_Lx} - b^*\,e^{-ik_Lx}\right)
& x \rightarrow -\infty \\
e^{ik_Rx} & x \rightarrow +\infty
\end{array}
\right.
\,.
\end{equation}
Similarly, we apply the transfer matrix (\ref{eq:trans}) to define $u_2$,
\begin{equation}
\label{eq:as2}
u_2(x) \sim {\cal A}_2
\left\{
\begin{array}{r@{\quad:\quad}l}
e^{-ik_Lx} & x \rightarrow -\infty\\
a\,e^{-ik_Rx} + b\,e^{ik_Rx}
& x \rightarrow +\infty
\end{array}
\right.
\,.
\end{equation}
We normalize the $A_1^\mathrm{in}$ and $A_2^\mathrm{in}$ modes
according to the scalar product (\ref{eq:scalar}), adopting the procedure
\cite{LL3} for normalizing Schr\"odinger waves in the continuous part
of the spectrum, which also shows that the $A_1^\mathrm{in}$ and
$A_2^\mathrm{in}$ are orthogonal to each other. We find
\begin{equation}
{\cal A}_1 = \frac{Z_L}{4\omega Z_R^2\,|a|^2}\,,\quad
{\cal A}_2 = \frac{1}{4\omega Z_R\,|a|^2}\,.
\end{equation}
In this way we have defined the two possible incident modes for
each frequency component of light in our effectively one-dimensional
situation. Consider the outgoing modes.
They are simply the incident modes traveling backwards,
\begin{equation}
A_1^\mathrm{out} = u_2^*(x)\,e^{-i\omega t} \,,\quad
A_2^\mathrm{out} = u_1^*(x)\,e^{-i\omega t} \,,
\end{equation}
forming an orthonormal set of modes as well.
Since the wave equation (\ref{eq:wave}) is of second order,
each set of modes establishes a basis.
Consequently, the outgoing modes are a superposition of the
ingoing ones,
\begin{equation}
\left(
    \begin{array}{c}
     A_1^\mathrm{out}  \\
     A_2^\mathrm{out}
    \end{array}
\right)
=
\underline{B}
\left(
    \begin{array}{c}
     A_1^\mathrm{in}  \\
     A_2^\mathrm{in}
    \end{array}
\right)\,,
\end{equation}
with the constant matrix $\underline{B}$, the scattering matrix
of the beam splitter.
We use the asymptotics (\ref{eq:as1}) and (\ref{eq:as2}) and the relation
(\ref{eq:det}) to determine the coefficients of  $\underline{B}$,
\begin{equation}
\label{eq:mediab} \underline{B} = \frac{1}{a} \left(
    \begin{array}{cc}
      \sqrt{Z_R/Z_L} & -b \\
      b^* & \sqrt{Z_R/Z_L}
    \end{array}
\right)
=
\frac{1}{a}
\left(
    \begin{array}{cc}
      \sqrt{|a|^2 - |b|^2} & -b \\
      b^* & \sqrt{|a|^2 - |b|^2}
    \end{array}
\right) \,.
\end{equation}
The scattering matrix is unitary
\begin{equation}
\label{eq:unib}
\underline{B}^{-1} =\underline{B}^\dagger \,.
\end{equation}
Therefore, as a consequence of the mode expansion (\ref{eq:modeex}),
the mode operators are transformed in precisely the same way as the
mode functions
\begin{equation}
\left(
    \begin{array}{c}
     \hat{a}_1^\mathrm{out}  \\
     \hat{a}_2^\mathrm{out}
    \end{array}
\right)
=
\underline{B}
\left(
    \begin{array}{c}
     \hat{a}_1^\mathrm{in}  \\
     \hat{a}_2^\mathrm{in}
    \end{array}
\right)\,.
\end{equation}
To summarize this section,
we may employ two alternative mode expansions (\ref{eq:modeex})
of the electromagnetic field, expansions in terms of incident
or of outgoing modes, with the scattering matrix as mediator.
The two sets of modes are adapted to two distinct physical situations
--- the incident modes refer to quantum light that enters the dielectric
structure from outside, whereas the outgoing modes are the ones
that leave the structure.

\section{Quantum-optical networks}

The scattering matrix completely characterizes a perfect piece of
dielectric structure in the regime of far-field optics,
describing how incident light beams are transformed into outgoing
beams. In one spatial dimension, two monochromatic modes
interfere to produce two emerging modes. In general, and
certainly in the three-dimensional real world, infinitely many
incident modes give rise to equally many outgoing modes. In
experimental quantum optics, one often tries to operate with as
few modes as possible. Much care is spent on aligning the
equipment to make sure that most of the quantum light of interest
is indeed captured in a few well-controlled modes. On the other
hand, one can construct, in a controlled way, optical networks,
also called multiports \cite{Walker},
from the basic building blocks such as
beam splitters and mirrors \cite{Mattle,Reck,Torma1,Torma3,Welsch},
networks with interesting quantum properties, in particular
in the limit when many elements are involved
\cite{Torma2,Torma4}.
Furthermore, we could add
phase-conjugating mirrors \cite{Shen} or parametric amplifiers
\cite{Shen} to our catalogue of simple optical instruments,
although they are experimentally less simple than dielectric
structures. Parametric amplifiers \cite{Shen} and
phase-conjugating mirrors \cite{Shen} are active devices --- they
require a source of energy, mostly light of a higher frequency.
Yet these active devices share a key property with the passive
instruments --- they are linear devices in the sense that the
input modes are linear transformations of the output modes.
However, this linear transformation may involve the Hermitian
conjugated mode operators. A phase-conjugating mirror, for
example, produces the complex-conjugated image of the incident
wave front $A_k$, which, in quantum optics, is associated with
the Hermitian conjugated mode operator $\hat{a}_k^\dagger$, the
creation operator.

\subsection{Linear mode transformations}

Assume that the set of mode operators $\hat{a}_k$ and their Hermitian
conjugates,  $\hat{a}_k^\dagger$, describing the incident quantum light,
is turned into the operators $\hat{a}_k'$ and  $\hat{a}_k'^\dagger$
of the outgoing modes, by the linear transformation
\begin{equation}
\label{eq:linear}
\left(
    \begin{array}{c}
     \hat{a}_k'  \\
     \hat{a}_k'^\dagger
    \end{array}
\right)
=
\underline{S}
\left(
    \begin{array}{c}
     \hat{a}_k  \\
     \hat{a}_k^\dagger
    \end{array}
\right)\,.
\end{equation}
The columns $(\hat{a}_k, \hat{a}_k^\dagger)^T$ and
$(\hat{a}_k', \hat{a}_k'^\dagger)^T$
refer to the total set of mode operators involved in the transformation.
We require that the operators of both the incident and the emerging
modes are indeed proper annihilation and creation operators,
subject to the commutation relations (\ref{eq:comm}),
written in matrix notation as
\begin{equation}
\label{eq:commatrix}
\big[\left(
    \begin{array}{c}
     \hat{a}_k  \\
     \hat{a}_k^\dagger
    \end{array}
\right),
(\hat{a}_{k'}^\dagger, \hat{a}_{k'})
\big]
=
\left(
    \begin{array}{cc}
      {[\hat{a}_k,  \hat{a}_{k'}^\dagger]} &
      {[\hat{a}_k,  \hat{a}_{k'}]} \\
      {[\hat{a}_k^\dagger,  \hat{a}_{k'}^\dagger]}  &
      {[\hat{a}_k^\dagger,  \hat{a}_{k'}]}
    \end{array}
\right)
= \underline{G}
\end{equation}
with
\begin{equation}
\underline{G} =
\left(
    \begin{array}{cc}
      \mathds{1} & 0 \\
       0 & -\mathds{1}
    \end{array}
\right)
\,.
\end{equation}
We substitute the mode transformation (\ref{eq:linear}) and its Hermitian
conjugate into the equivalent relation for $\hat{a}_k'$ and
$\hat{a}_k'^\dagger$, and find
\begin{equation}
\label{eq:qu}
\underline{S}\,\underline{G}\,\underline{S}^\dagger = \underline{G}\,.
\end{equation}
Such transformations are called quasi-unitary \cite{Cornwell}. In the
classical mechanics of a many-particle system \cite{LL1}, they
are called linear canonical transformations, because they
preserve the canonical Poisson brackets \cite{LL1}. Equation
(\ref{eq:qu}) implies that $\underline{G}\,\underline{S}^\dagger
\underline{G}$ is the inverse of $\underline{S}$. Therefore,
$\underline{S}$ is a square matrix. We get for the determinant
\begin{equation}
\label{eq:jacobi} \left|\, \mathrm{det}\, \underline{S}
\,\right|^2 = 1 \,.
\end{equation}
Consequently, the Jacobian of the mode transformation has unity
modulus --- the phase-space volume is conserved, as we would
expect from canonical transformations according to Liouville's
theorem \cite{LL1}. Quantum mechanics requires that the number of
input modes is exactly the same as the number of output modes.
Modes that are ``empty'' contribute nevertheless to the quantum
properties of the device. They are not really empty, they are just in
the vacuum state. The vacuum noise behind a mirror matters
\cite{Leonhardt}, and so do the vacuum fluctuations of modes
prior to amplification.

When the optical instrument transforms annihilation operators into
annihilation operators without involving their Hermitian
conjugates, as it is the case for the one-dimensional dielectric
structures analyzed in Sec.\ 2 or for passive optical multiports in
general \cite{Mattle,Reck,Torma1,Torma3,Walker}, we get
\begin{equation}
\label{eq:unit} \underline{S} = \left(
    \begin{array}{cc}
      \underline{B} & 0 \\
       0 & {\underline{B}^*}
    \end{array}
\right) \,,\quad \underline{B}\,\underline{B}^\dagger =  \mathds{1}
\,.
\end{equation}
We see that the unitarity (\ref{eq:unib}) of the beam-splitter
matrix $\underline{B}$ is not a coincidence
--- it follows from the conservation of the commutation relation
(\ref{eq:comm}) in light scattering. As a consequence of the
unitarity of $\underline{B}$, the total number of photons is
conserved,
\begin{equation}
\label{eq:totalnumber}
\sum_k \hat{a}_k'^\dagger \hat{a}_k' =
\sum_k \hat{a}_k^\dagger \hat{a}_k
\,.
\end{equation}
Passive optical instruments conserve the total energy. Active
devices such as phase-conjugating mirrors \cite{Shen} or parametric
amplifiers \cite{Shen} combine annihilation and creation operators.
Consequently, the total number of photons is not conserved, in
general, indicating that active devices rely on external energy
sources.

\subsection{Quantum-state transformations}

The linear transformation (\ref{eq:linear}) of mode operators describes
the transformation of the incident quantum light into the emerging
quanta in a peculiar manner.
In the mode transformations, the quantum state of light is invariant,
but its relation to physical observables changes,
similar to the Heisenberg picture of quantum mechanics.
A given quantum superposition of photons, frozen in space and time,
is seen first as constituting the incident modes and then as leaving
in the outgoing modes.
The Schr\"odinger picture gives perhaps a more natural approach
to understanding the quantum effects of optical instruments on light.
Here the instrument changes the state of light, whereas the mode
operators remain invariant. In other words, in the Schr\"odinger
picture the incident and the emerging modes are the same and the
optical instrument operates like a black box on the quantum state of light.
This picture is especially suitable for analyzing laboratory situations
where a few well-controlled modes enter the instrument and leave it
in other equally-well-controlled modes, describing how the quantum
state is transformed. The Heisenberg and the Schr\"odinger picture
ought to agree on their quantum-mechanical predictions, on
expectation values, and this is how we deduce the quantum-state
transformations from the linear mode transformations (\ref{eq:linear})
\cite{LeoNeu}.

We describe the quantum state of light in terms of the density operator
(also called density matrix) \cite{Carmichael,Gardiner,LL5,Leonhardt}.
We require for any physical observable of light, for any
function of the mode operators, that the expectation values in
the two pictures agree,
\begin{equation}
\mathrm{tr}\{\hat{\rho}f(\hat{a}_k',\hat{a}_k'^\dagger)\} =
\mathrm{tr}\{\hat{\rho}'f(\hat{a}_k,\hat{a}_k^\dagger)\}
\,,
\end{equation}
where $\hat{\rho}'$ denotes the density operator of the outgoing light.
If we manage to determine a unitary evolution operator $\hat{B}$
with the property
\begin{equation}
\label{eq:bop}
\left(
    \begin{array}{c}
     \hat{a}_k'  \\
     \hat{a}_k'^\dagger
    \end{array}
\right)
=
\underline{S}
\left(
    \begin{array}{c}
     \hat{a}_k  \\
     \hat{a}_k^\dagger
    \end{array}
\right)
=
\hat{B}
\left(
    \begin{array}{c}
     \hat{a}_k  \\
     \hat{a}_k^\dagger
    \end{array}
\right)
\hat{B}^\dagger
\end{equation}
we get
\begin{equation}
\hat{\rho}' = \hat{B}^\dagger \hat{\rho}\,\hat{B}
\,.
\end{equation}
Consider the logarithm of $\underline{S}$ defined as a matrix
$\ln \underline{S}$ for which
\begin{equation}
\label{eq:log} \exp(\ln \underline{S}) = \sum_{n=0}^\infty
\frac{(\ln \underline{S})^n}{n!} = \underline{S} \,.
\end{equation}
The quasi-unitarity (\ref{eq:qu}) of the $\underline{S}$ matrix
implies that $-\ln\underline{S}=\ln(\underline{S}^{-1})=
\ln(\underline{G}\,\underline{S}^\dagger\underline{G})$.
Furthermore, since $\underline{G}^2=\mathds{1}$, we obtain from the
definition (\ref{eq:log}) of the matrix logarithm that
$\ln(\underline{G}\,\underline{S}^\dagger\underline{G})=
\underline{G}\,(\ln\underline{S})^\dagger\underline{G}$.
Consequently,
\begin{equation}
\label{eq:pdef} \underline{H} = -i\underline{G}\ln\underline{S}
\end{equation}
is a Hermitian matrix. We construct the operators \cite{LeoNeu}
\begin{equation}
\label{eq:ham} \hat{B} = \exp(i\hat{H}) \,,\quad \hat{H} =
\frac{1}{2}\,(\hat{a}_k^\dagger, \hat{a}_k)\, \underline{H} \left(
    \begin{array}{c}
     \hat{a}_k  \\
     \hat{a}_k^\dagger
    \end{array}
\right)
\,.
\end{equation}
Since $\hat{H}$ is Hermitian, $\hat{B}$ is unitary. We prove that
$\hat{B}$ does indeed act as an evolution operator with the
property (\ref{eq:bop}). Consider the power $\hat{B}^\eta$ for
real $\eta$. First we show that the differential equation in
$\eta$ of $\hat{B}^\eta \hat{a}_k \hat{B}^{\dagger\eta}$ is the
same as the differential equation for the transformed mode
operators with matrix $\underline{S}^\eta$. From the commutation
relations (\ref{eq:comm}) in matrix form (\ref{eq:commatrix})
follows
\begin{eqnarray}
\partial_\eta\, \hat{B}^\eta
\left(
    \begin{array}{c}
     \hat{a}_k  \\
     \hat{a}_k^\dagger
    \end{array}
\right) \hat{B}^{\dagger\eta} &=& i\exp(i\eta\hat{H})\,
\big[\hat{H}, \left(
    \begin{array}{c}
     \hat{a}_k  \\
     \hat{a}_k^\dagger
    \end{array}
\right) \big] \exp(-i\eta\hat{H})
\nonumber\\
&=&  i\exp(i\eta\hat{H})\,
 \underline{G}\,\underline{H} \left(
    \begin{array}{c}
     \hat{a}_k  \\
     \hat{a}_k^\dagger
    \end{array}
\right)
\exp(-i\eta\hat{H})
\nonumber\\
&=& i\underline{G}\,\underline{H}\,\hat{B}^\eta \left(
    \begin{array}{c}
     \hat{a}_k  \\
     \hat{a}_k^\dagger
    \end{array}
\right) \hat{B}^{\dagger\eta} \,,
\end{eqnarray}
which indeed agrees with the differential equation
\begin{equation}
\partial_\eta\,
\underline{S}^\eta \left(
    \begin{array}{c}
     \hat{a}_k  \\
     \hat{a}_k^\dagger
    \end{array}
\right) = i\underline{G}\,\underline{H}\,\underline{S}^\eta \left(
    \begin{array}{c}
     \hat{a}_k  \\
     \hat{a}_k^\dagger
    \end{array}
\right)
\end{equation}
for $\underline{H}$ given in terms (\ref{eq:pdef}) of the matrix
logarithm. Since at $\eta=0$ the mode operators are not
transformed, the initial $\hat{B}^\eta \hat{a}_k
\hat{B}^{\dagger\eta}$ trivially agrees with the effect of
$\underline{S}^0=\mathds{1}$. Therefore, $\hat{B}^\eta(\hat{a}_k,
\hat{a}_k^\dagger)^T\hat{B}^{\dagger\eta}$ gives
$\underline{S}^\eta(\hat{a}_k, \hat{a}_k^\dagger)^T$ all the way
up to $\eta=1$, thus proving the relation (\ref{eq:bop}).

The $\hat{H}$ operator plays the role of the effective Hamiltonian
for the quantum-optical network \cite{Torma1,Torma3}, generating
the linear mode transformation (\ref{eq:linear}) in the Heisenberg
picture. This Hamiltonian depends on the logarithm of the
transformation matrix $\underline{S}$, which is a multivalued
function with infinitely many branches. So there are
many equivalent ways to design an optical network with a
particular input-output relation and there are also many ways
to assemble it from the
basic building blocks \cite{Reck,Torma1,Torma3}, from beam
splitters and parametric amplifiers.

\subsection{Examples}

The two prime examples of simple optical instruments are the beam
splitter and the parametric amplifier. We describe the beam
splitter by the unitary scattering matrix $\underline{B}$. For
simplicity, let us assume that $\underline{B}$ is a real rotation
matrix instead of the general $2 \times 2$ unitary matrix,
\begin{equation}
\label{eq:rot} \underline{B} = \left(
    \begin{array}{cc}
      \cos\varphi & \sin\varphi \\
       -\sin\varphi & \cos\varphi
    \end{array}
\right)
\end{equation}
with the rotation angle $\varphi$. For example, the matrix
(\ref{eq:rot}) may describe a polarizing beam splitter where an
incident light beam is separated into two linear polarizations
with angles $\varphi$ and $\varphi+\pi$. We show in Sec.\ 4 that
the simple rotation matrix (\ref{eq:rot}) contains the essence of
all two-mode beam splitters. Here we derive the Hamiltonian for
the linear mode transformation (\ref{eq:linear}) described by the
beam-splitter matrix (\ref{eq:rot}). Consider the matrix
\begin{equation}
\underline{I} =
    \left(
    \begin{array}{cccc}
      0&1&0&0\\-1&0&0&0\\0&0&0&1\\0&0&-1&0
    \end{array}
\right) \,.
\end{equation}
Since $\underline{I}^2 = -\mathds{1}$ we get
\begin{eqnarray}
\exp(\varphi\underline{I}) &=& \sum_{k=0}^\infty
\frac{\varphi^{2k}}{(2k)!}\,(-1)^k \mathds{1} + \sum_{k=0}^\infty
\frac{\varphi^{2k+1}}{(2k+1)!}\,(-1)^k \underline{I} \nonumber\\
&=& (\cos\varphi)\mathds{1} + (\sin\varphi) \underline{I} \,.
\end{eqnarray}
Consequently, $\varphi\underline{I}$ is the logarithm of the
$\underline{S}$ matrix with $\underline{S}$ given by Eq.\
(\ref{eq:unit}) and the beam-splitter matrix (\ref{eq:rot}).
Therefore, the effective Hamiltonian
(\ref{eq:ham}) of the beam splitter is
\begin{equation}
\label{eq:hbeam} \hat{H} =
i\varphi\left(\hat{a}_1^\dagger\hat{a}_2 -
\hat{a}_2^\dagger\hat{a}_1\right) \,.
\end{equation}
The Hamiltonian indicates that photons from mode 1 are
annihilated and converted into photons of mode 2, and vice versa,
which is just what we expect from a beam splitter.
As we know, the total
number of photons is conserved for passive optical instruments.

Let us turn to active devices that, by definition, mix
annihilation and creation operators.
A simple example of an active device
is characterized by the $\underline{S}$ matrix
\begin{equation}
\label{eq:lorentz} \underline{S} =
    \left(
    \begin{array}{cccc}
      \cosh\zeta&0&0&\sinh\zeta\\ 0&\cosh\zeta&\sinh\zeta&0\\
      0&\sinh\zeta&\cosh\zeta&0\\ \sinh\zeta&0&0&\cosh\zeta
    \end{array}
\right)
\end{equation}
with the real parameter $\zeta$. One easily verifies that the
matrix (\ref{eq:lorentz}) indeed satisfies the quasi-unitarity relation
(\ref{eq:qu}) and thus qualifies for the $\underline{S}$ matrix
of an optical instrument. In fact, the matrix describes a
parametric amplifier \cite{Shen} or a phase-conjugating mirror
\cite{Shen}, see Sec.\ 6. We derive the effective Hamiltonian.
Consider the matrix
\begin{equation}
\underline{E} =
    \left(
    \begin{array}{cccc}
      0&0&0&1\\0&0&1&0\\0&1&0&0\\1&0&0&0
    \end{array}
\right) \,.
\end{equation}
Since $\underline{E}^2 = \mathds{1}$ we get
\begin{equation}
\exp(i\zeta\underline{E}) = (\cosh\zeta)\mathds{1} + (\sinh\zeta)
\underline{E} = \underline{S} \,.
\end{equation}
Consequently, the effective Hamiltonian (\ref{eq:ham}) is
\begin{equation}
\label{eq:hamp} \hat{H} = i\zeta\left(
\hat{a}_1^\dagger\hat{a}_2^\dagger-\hat{a}_1\hat{a}_2 \right) \,.
\end{equation}
The Hamiltonian indicates that two photons are simultaneously
created or annihilated. The pump process of the amplifier,
accounted for in the parameter $\zeta$, must provide the energy
source of the photon-pair production or the reservoir for
annihilation. We could represent $\zeta$ as $\gamma t /2$ where
$t$ denotes the amplification time and $\gamma$ the differential
gain that depends on the performance of the pump. In our simple
model (\ref{eq:lorentz}) the pump is assumed to be classical and
to remain essentially unchanged, giving rise to a constant rate
$\gamma$ during the amplification.

\subsection{Wigner function}

Quasiprobability distributions \cite{BK,Leonhardt,Schleich}
are frequently
used in quantum optics to draw intuitive pictures of the quantum
fluctuations of light, for computational advantages and to give a
precise meaning to the notion of non-classical light
\cite{Leonhardt}. Quasiprobability distributions are functions of
the classical quadratures $q$ and $p$ that behave in many ways
like classical probability densities. However, since the
quadrature operators $\hat{q}$ and $\hat{p}$ do not commute,
since $\hat{q}$ and $\hat{p}$ cannot be measured simultaneously
{\it and} precisely, the quasiprobability distributions must not
represent perfect phase-space densities. For example, they may
appear to describe negative probabilities or they may become
mathematically ill behaved \cite{Leonhardt}. In quantum optics,
the most prominent quasiprobability distributions are the $P$
function, the $Q$ function and the Wigner function,
see for example Ref.\ \cite{Leonhardt}.
The $P$ function allows us to express, by the
optical equivalence theorem \cite{Glauber,Leonhardt,Sudarshan},
any quantum state as a quasi-ensemble of coherent states (of
classical light waves) \cite{Leonhardt,Loudon,MandelWolf}. If the
$P$ function is non-negative and well-behaved the light is
said to be classical.
Otherwise the light is non-classical in the sense that
it cannot be understood as partially coherent classical light. The
$Q$ function is proportional to the expectation value of the
density matrix in coherent states \cite{Leonhardt}. The $Q$
function appears as the genuine probability distribution in
simultaneous measurements of position and momentum quadratures
\cite{Leonhardt,WalkerCarroll,Walker}. However, since $q$ and $p$
cannot be measured both simultaneously and precisely, the $Q$
function contains some extra quantum noise that is difficult to
remove from the true quantum state by deconvolutions
\cite{LPSim,Leonhardt}.

The Wigner function
 \cite{Leonhardt,Moyal,Schleich,Tatarskii,Wigner}
is probably best
suited to describe the quantum effects of simple optical
instruments. The Wigner function $W(q,p)$ of a single mode of
light is the inverse Fourier transform of the characteristic
function $\widetilde{W}(u,v)$ \cite{Leonhardt}
\begin{eqnarray}
\label{eq:def} W(q,p) &=& \frac{1}{(2\pi)^2}
\int_{-\infty}^{+\infty} \int_{-\infty}^{+\infty}
\widetilde{W}(u,v)\, \exp(iuq+ivp)\,
\mathrm{d}u\,\mathrm{d}v \,, \nonumber\\
\widetilde{W}(u,v) &=&
\mathrm{tr}\{\hat{\rho}\exp(-iu\hat{q}-iv\hat{p})\} \,.
\end{eqnarray}
We obtain \cite{Leonhardt} in terms of the $q$ quadrature
eigenstates (position eigenstates)
\begin{equation}
\label{eq:wigner} W(q,p) = \frac{1}{2\pi}
\int_{-\infty}^{+\infty} \exp(ipx) \langle
q-x/2\,|\,\hat{\rho}\,|\,q+x/2\rangle\,\mathrm{d}x \,.
\end{equation}
The Wigner function is real and normalized to unity for any
proper density operator \cite{Leonhardt}. Quantum expectation
values can be computed via the overlap formula \cite{Leonhardt}
\begin{equation}
\label{eq:overlap} \mathrm{tr}\{\hat{F}_1\hat{F_2}\} = 2\pi
\int_{-\infty}^{+\infty} \int_{-\infty}^{+\infty}
W_1(q,p)\,W_2(q,p)\, \mathrm{d}q\,\mathrm{d}p \,,
\end{equation}
where $W_1$ and $W_2$ are the Wigner transforms with $\hat{\rho}$
in formula (\ref{eq:wigner}) replaced by $\hat{F}_1$ and
$\hat{F}_2$, respectively. The Wigner function gives a faithful
and frequently quite intuitive image of the quantum state of a
single mode of light. Optical homodyne tomography has been
applied to reconstruct the Wigner function from homodyne
measurements \cite{Breitenbach,Leonhardt,Lvovsky,Smithey,Welsch}.
The marginal distributions of the Wigner function agree with
the correct quadrature histograms
with respect to an arbitrary phase shift.
This tomographic principle underlies optical homodyne tomography
and it also defines the Wigner function
uniquely \cite{Bertrand,Leonhardt}. The Wigner
function represents a fairly good compromise between the abstract density
operator of quantum mechanics and the phase-space density of
classical statistical mechanics, but the Wigner function may
exhibit negative ``probabilities'' in small phase-space regions
\cite{Leonhardt}. Such features are quite subtle and have been
observed only recently in quantum light \cite{Bertet,Lvovsky}.

Here we use the Wigner function to describe the quantum effects
generated by simple optical instruments that are subject to the
linear mode transformations (\ref{eq:linear}). We extend the
definition (\ref{eq:def}) of the Wigner function to a multitude
of light modes characterized by the classical amplitudes
\begin{equation}
\alpha_k = \frac{1}{\sqrt{2}} (q_k+ip_k) \,.
\end{equation}
We represent $u\hat{q}_k+v\hat{p}_k$ in the definition
(\ref{eq:def}) of the characteristic function as
$\beta_k^*\hat{a}_k+\beta_k\hat{a}_k^\dagger$ with
$\beta_k=(u_k+iv_k)/\sqrt{2}$. We see that
\begin{eqnarray}
\widetilde{W}'(\beta_k,\beta_k^*) &=&
\mathrm{tr}\left\{\hat{\rho}'\, \exp\left(-i\sum_k
(\beta_k^*\hat{a}_k+\beta_k\hat{a}_k^\dagger)\right)\right\}
\nonumber\\
&=& \mathrm{tr}\left\{\hat{\rho}\, \exp\left(-i\sum_k
(\beta_k^*\hat{a}_k'+\beta_k\hat{a}_k'^\dagger)\right)\right\}
\nonumber\\
&=& \widetilde{W}(\beta_k',\beta_k'^*)
\end{eqnarray}
with
\begin{equation}
\left(
    \begin{array}{c}
     \beta_k'  \\
     \beta_k'^*
    \end{array}
\right) = \underline{S}^{-1} \left(
    \begin{array}{c}
     \beta_k  \\
     \beta_k^*
    \end{array}
\right)\,.
\end{equation}
To obtain the Wigner function (\ref{eq:def}) of the emerging
multi-mode light we represent $uq_k+vp_k$ as
$\beta_k^*\alpha_k+\beta_k\alpha_k^*$ in the inverse Fourier
transformation of the characteristic function and we use
$\beta_k$ and $\beta_k^*$ as the integration variables. Then we
perform a variable transformation from $\beta_k$ and $\beta_k^*$
to $\beta_k'$ and $\beta_k'^*$. The Jacobian (\ref{eq:jacobi}) of
this transformation has unity modulus, and we get the result
\begin{equation}
\label{eq:wignertrans} W'(\alpha_k,\alpha_k^*) =
W(\alpha_k',\alpha_k'^*) \,,\quad \left(
    \begin{array}{c}
     \alpha_k'  \\
     \alpha_k'^*
    \end{array}
\right) = \underline{S}^{-1} \left(
    \begin{array}{c}
     \alpha_k  \\
     \alpha_k^*
    \end{array}
\right)\,.
\end{equation}
Simple optical instruments transform the Wigner function $W$ of
the incident quantum light as if $W$ were a classical probability
distribution of the mode amplitudes. This property uniquely
distinguishes \cite{EkertKnight} the Wigner function for general
quasi-unitary transformations involving Hermitian conjugated mode
operators, {\it i.e.}\ for active optical instruments
\cite{LeoSU11}. Passive instruments such as optical multiports
\cite{Reck,Torma1,Torma3} transform also the $P$ function and the
$Q$ function like classical probability distributions
\cite{LeoSU2}.

\section{Beam splitter}

\begin{figure}
\begin{center}
\includegraphics[width=10cm]{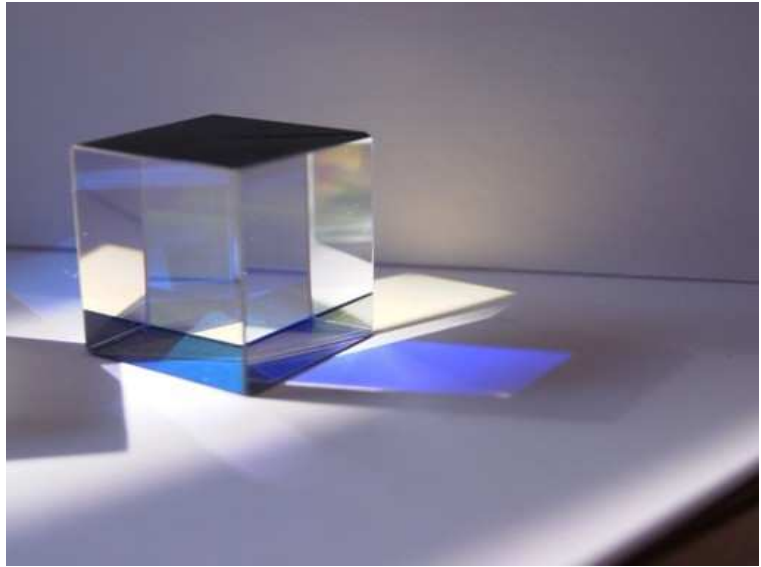}\vspace*{1cm}
\includegraphics[width=7cm]{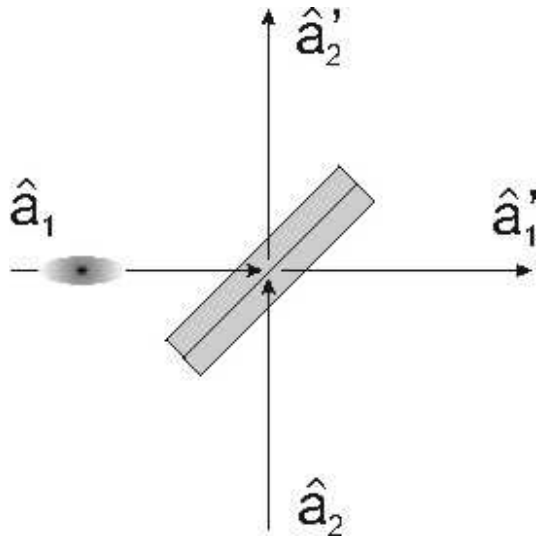}
\end{center}
\caption{\label{fig:beamsplitter}Beam splitter.
The picture above shows a polarizing beam splitter.
(Courtesy of Oliver Gl\"ockl and Natasha Korolkova.)
The picture below schematically illustrates
the quantum theory of the beam splitter.
Two incident light modes, represented by the
Bose mode operators $\hat{a}_1$ and $\hat{a}_2$
interfere to produce two emerging modes
with operators $\hat{a}_1'$ and $\hat{a}_2'$.
Even if only one incident beam is split
the vacuum noise of the second incident
mode behind the mirror plays an important
role in the quantum optics of the beam splitter.}
\end{figure}

The archetype of passive optical instruments is the beam
splitter, usually an innocent-looking cube of glass in laboratory
experiments, see Fig.\ \ref{fig:beamsplitter}.
Light incident at the front of the cube is split into
two beams, and so is light incident at the back. Both modes may
interfere. In Sec.\ 2 we studied the theoretically simplest
example of a beam splitter, a one-dimensional dielectric
structure. Polarizers, where the two polarization modes of light
are mixed, are also essentially beam splitters,
and so are simple passive interferometers.
The theory of this passive four-port device has been developed in Refs.\
\cite{Allen,Aharonov,Brunner,Campos,Huttner,Janszky1,Janszky2,Janszky3,Lai,LeoSU2,Luis,OHM,PaulSplitter,PaulReview,Perinova,Prasad,vanderPlank,Zeilinger}.
Here we follow mostly Refs.\ \cite{Campos,LeoSU2}.
More complicated
optical multiports \cite{Reck,Torma1,Torma3}, where a multitude
of beams interfere to produce the same number of outgoing modes,
can be constructed from beam splitters and mirrors \cite{Reck}.
But already the simple beam splitter, combined with good
photodetectors and single-photon sources, is quite capable of
demonstrating some fundamental aspects of the wave-particle
duality of light.

\subsection{Matrix structure}

The beam splitter is completely characterized by a unitary
$2\times2$ matrix $\underline{B}$ that describes how the device
transforms the incident modes into the outgoing modes in the
Heisenberg picture,
\begin{equation}
\label{eq:bs} \left(
    \begin{array}{c}
     \hat{a}_1'  \\
     \hat{a}_2'
    \end{array}
\right) = \underline{B} \left(
    \begin{array}{c}
     \hat{a}_1  \\
     \hat{a}_2
    \end{array}
\right)\,,\quad \underline{B} = \left(
    \begin{array}{cc}
      B_{11} & B_{12} \\
      B_{21} & B_{22}
    \end{array}
\right)\,.
\end{equation}
The beam-splitter matrix is unitary, in order to preserve the
Bose commutation relations between the mode operators.
Explicitly, the matrix elements must obey
\begin{equation}
|\,B_{11}\,|^2+|\,B_{12}\,|^2=1\,,\quad
|\,B_{21}\,|^2+|\,B_{22}\,|^2=1\,,\quad
B_{11}^*B_{21}+B_{12}^*B_{22}=0\,.
\end{equation}
The general solution of these equations is
\begin{equation}
\label{eq:bmatrix} \underline{B} = e^{i\Lambda/2}\left(
    \begin{array}{cc}
      \cos(\Theta/2)\,e^{i(\Psi+\Phi)/2} &
      \sin(\Theta/2)\,e^{i(\Psi-\Phi)/2}  \\
      -\sin(\Theta/2)\,e^{i(\Phi-\Psi)/2}  &
      \cos(\Theta/2)\,e^{-i(\Psi+\Phi)/2}
    \end{array}
\right)
\end{equation}
with the real parameters $\Phi$, $\Theta$, $\Psi$ and $\Lambda$.
The one-dimensional dielectric structure with matrix
(\ref{eq:mediab}) represents the special case where $\Phi=-\Psi$.
We express the general beam-splitter matrix (\ref{eq:bmatrix}) as
the product
\begin{equation}
\label{eq:u2} \underline{B} = e^{i\Lambda/2} \left(
    \begin{array}{cc}
      e^{i\Psi/2} & 0 \\
      0 & e^{-i\Psi/2}
    \end{array}\right) \left(
    \begin{array}{cc}
      \cos(\Theta/2) & \sin(\Theta/2) \\
      -\sin(\Theta/2) & \cos(\Theta/2)
    \end{array}
\right)\left(
    \begin{array}{cc}
      e^{i\Phi/2} & 0 \\
      0 & e^{-i\Phi/2}
    \end{array}\right)\,.
\end{equation}
The beam splitter acts in four steps. The incident modes gain a
relative phase of $\Phi$, the modes are optically mixed with the
mixing angle $\Theta/2$, and the outgoing modes attain the
relative phase $\Psi$ and the overall phase $\Lambda/2$. We could
incorporate the phases into the definitions of the incident and
the outgoing modes. The rotation matrix would remain as the key
feature of the beam splitter. The reflectivity $\varrho$ is
characterized by $-\sin(\Theta/2)$ (the sign is unimportant
though) while the transmissivity $\tau$ is given by
$\cos(\Theta/2)$, which implies $\tau^2+\varrho^2=1$. The beam
splitter has been assumed to be perfectly lossless --- if a photon
is not transmitted it must be reflected. We show in Sec.\ 5 how,
in principle, absorption can be included and that the beam
splitter itself serves as a convenient model of an absorber.

\subsection{Quantum Stokes parameters}

\begin{figure}
\begin{center}
\includegraphics[width=7cm]{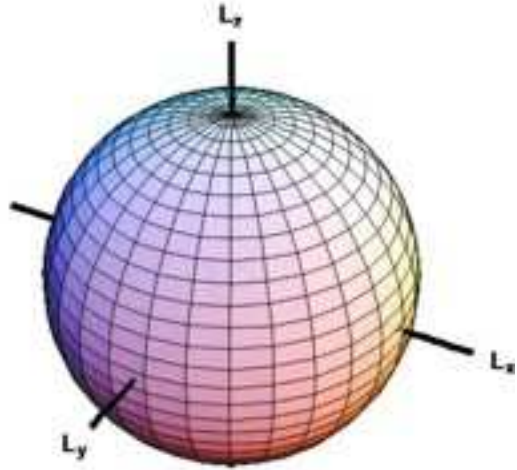}
\end{center}
\caption{\label{fig:sphere}Poincar\'e sphere.
The classical polarization of light is characterized by the
Stokes parameters that lie on the Poincar\'e sphere.
In quantum optics, the Jordan-Schwinger operators (\ref{eq:js})
play the role of the quantum Stokes parameters
and their quantum statistics characterizes the
polarization state.
The parameter $\hat{L}_t$,
the conserved total photon number,
corresponds to the radius of the Poincar\'e sphere.
The polarizer of Fig.\ \ref{fig:beamsplitter}
rotates the quantum Stokes parameters.
In general, we can characterize the two incident modes
of any beam splitter
by their Jordan-Schwinger operators as well,
and the beam splitter performs
a rotation on the generalized Poincar\'e sphere.}
\end{figure}

The classical polarization of a light beam is usually described
using the Stokes parameters \cite{BornWolf}. Given the complex
amplitudes $a_1$ and $a_2$ of the two polarization modes, the
three Stokes parameters are proportional to the corresponding
expectation values of the Pauli matrices
\begin{equation}
\label{eq:pauli}
\sigma_x = \left(
    \begin{array}{cc}
      0 & 1 \\
      1 & 0
    \end{array}
\right)\,,\quad\sigma_y = \left(
    \begin{array}{cc}
      0 & -i \\
      i & 0
    \end{array}
\right)\,,\quad\sigma_z = \left(
    \begin{array}{cc}
      1 & 0 \\
      0 & -1
    \end{array}
\right)\,.
\end{equation}
The Stokes parameters lie on a sphere, the Poincar\'e sphere
\cite{BornWolf}, also called the Bloch sphere in quantum
mechanics \cite{MandelWolf}, see Fig.\ \ref{fig:sphere}.
In quantum  optics, we describe the
polarization, the spinor part of the angular momentum of light,
in terms of the quantum Stokes parameters \cite{Korolkova}
\begin{eqnarray}
\hat{L}_t &=& \frac{1}{2}\, (\hat{a}_1^\dagger,\,
\hat{a}_2^\dagger)\,\, \mathds{1}\, \left(
    \begin{array}{c}
     \hat{a}_1  \\
     \hat{a}_2
    \end{array}
\right) = \frac{1}{2}\,\left(\hat{a}_1^\dagger\hat{a}_1 +
\hat{a}_2^\dagger\hat{a}_2\right)\,, \nonumber\\
\hat{L}_x &=& \frac{1}{2}\, (\hat{a}_1^\dagger,\,
\hat{a}_2^\dagger)\, \sigma_x \left(
    \begin{array}{c}
     \hat{a}_1  \\
     \hat{a}_2
    \end{array}
\right) = \frac{1}{2}\,\left(\hat{a}_1^\dagger\hat{a}_2 +
\hat{a}_2^\dagger\hat{a}_1\right)\,, \nonumber\\
\hat{L}_y &=& \frac{1}{2}\, (\hat{a}_1^\dagger,\,
\hat{a}_2^\dagger)\, \sigma_y \left(
    \begin{array}{c}
     \hat{a}_1  \\
     \hat{a}_2
    \end{array}
\right) = \frac{i}{2}\,\left(\hat{a}_2^\dagger\hat{a}_1 -
\hat{a}_1^\dagger\hat{a}_2\right)\,, \nonumber\\
\hat{L}_z &=& \frac{1}{2}\, (\hat{a}_1^\dagger,\,
\hat{a}_2^\dagger)\, \sigma_z \left(
    \begin{array}{c}
     \hat{a}_1  \\
     \hat{a}_2
    \end{array}
\right) = \frac{1}{2}\,\left(\hat{a}_1^\dagger\hat{a}_1 -
\hat{a}_2^\dagger\hat{a}_2\right)\,,
\label{eq:js}
\end{eqnarray}
that obey the commutation relations of angular-momentum operators
\begin{equation}
\label{eq:amc} [\hat{L}_x,\,\hat{L}_y]=i\hat{L}_z \,,\quad
[\hat{L}_y,\,\hat{L}_z]=i\hat{L}_x \,,\quad
[\hat{L}_z,\,\hat{L}_x]=i\hat{L}_y \,.
\end{equation}
Equation (\ref{eq:js}) is called the Jordan-Schwinger
representation \cite{Jordan,Schwinger} of the angular momentum in
terms of two Bose operators. The representation serves as the
starting point for the quantum theory of polarized or partially
polarized light \cite{Korolkova,Lehner}. The operator $\hat{L}_t$
commutes with all others and serves to represent
the squared total angular momentum,
\begin{equation}
\label{eq:totall} \hat{L}_x^2 + \hat{L}_y^2 + \hat{L}_z^2 =
\hat{L}_t(\hat{L}_t + 1) \,.
\end{equation}
The commutation
relations (\ref{eq:amc}) give rise to uncertainty relations
between the quantum Stokes parameters. Polarization squeezing
\cite{Heersink} occurs when the statistical fluctuations of one
Stokes parameter are below the minimum-uncertainty limit.
This quantum-noise reduction of the polarization of light
has been applied to
observe macroscopic spin-squeezing effects in atomic vapors
\cite{Hald,Julsgaard}.

The Jordan-Schwinger representation (\ref{eq:js}) serves not only
to characterize the polarization of quantum light, the
representation provides also the theoretical tools to describe
the effect of polarizers or of any beam splitter in general. We
obtain from the commutation relations (\ref{eq:amc})
\begin{eqnarray}
\exp(-i\Phi\hat{L}_z) \left(
    \begin{array}{c}
     \hat{L}_x \\
     \hat{L}_y \\
     \hat{L}_z
    \end{array}
\right) \exp(i\Phi\hat{L}_z) &=& \left(
    \begin{array}{ccc}
      \cos\Phi & \sin\Phi & 0\\
      -\sin\Phi & \cos\Phi & 0\\
      0 & 0 & 1
    \end{array}
\right) \left(
    \begin{array}{c}
     \hat{L}_x \\
     \hat{L}_y \\
     \hat{L}_z
    \end{array}
\right)\,, \nonumber\\
\exp(-i\Theta\hat{L}_y) \left(
    \begin{array}{c}
     \hat{L}_x \\
     \hat{L}_y \\
     \hat{L}_z
    \end{array}
\right) \exp(i\Theta\hat{L}_y) &=& \left(
    \begin{array}{ccc}
      \cos\Theta & 0 & \sin\Theta\\
      0 & 1 & 0\\
      -\sin\Theta & 0 & \cos\Theta
    \end{array}
\right) \left(
    \begin{array}{c}
     \hat{L}_x \\
     \hat{L}_y \\
     \hat{L}_z
    \end{array}
\right)\,, \nonumber\\
\exp(-i\Omega\hat{L}_x) \left(
    \begin{array}{c}
     \hat{L}_x \\
     \hat{L}_y \\
     \hat{L}_z
    \end{array}
\right) \exp(i\Omega\hat{L}_x) &=& \left(
    \begin{array}{ccc}
      1 & 0 & 0\\
      0 & \cos\Omega & \sin\Omega\\
      0 & -\sin\Omega & \cos\Omega
    \end{array}
\right) \left(
    \begin{array}{c}
     \hat{L}_x \\
     \hat{L}_y \\
     \hat{L}_z
    \end{array}
\right)\,,
\label{eq:poincarerot}
\end{eqnarray}
as one easily verifies by differentiation with respect to the
parameters. Therefore, the exponential Jordan-Schwinger operators
describe rotations on the Poincare sphere, generated by
polarizers. We note that the angles $\Phi$, $\Theta$, $\Psi$ in
the complex matrix representation (\ref{eq:bmatrix}) are the
Euler angles of an arbitrary rotation in three-dimensional space
\cite{LL1}. We obtain for the mode operators
\begin{eqnarray}
\exp(-i\Phi\hat{L}_z) \left(
    \begin{array}{c}
     \hat{a}_1 \\
     \hat{a}_2
    \end{array}
\right) \exp(i\Phi\hat{L}_z) &=& \left(
    \begin{array}{cc}
      e^{i\Phi/2} & 0\\
      0 & e^{-i\Phi/2}\\
    \end{array}
\right) \left(
    \begin{array}{c}
     \hat{a}_1 \\
     \hat{a}_2
    \end{array}
\right) \,, \nonumber\\
\exp(-i\Theta\hat{L}_y) \left(
    \begin{array}{c}
     \hat{a}_1 \\
     \hat{a}_2
    \end{array}
\right) \exp(i\Theta\hat{L}_y) &=& \left(
    \begin{array}{cc}
      \cos(\Theta/2) & \sin(\Theta/2)\\
      -\sin(\Theta/2) & \cos(\Theta/2)\\
    \end{array}
\right) \left(
    \begin{array}{c}
     \hat{a}_1 \\
     \hat{a}_2
    \end{array}
\right) \,, \nonumber\\
\exp(-i\Omega\hat{L}_x) \left(
    \begin{array}{c}
     \hat{a}_1 \\
     \hat{a}_2
    \end{array}
\right) \exp(i\Omega\hat{L}_x) &=& \left(
    \begin{array}{cc}
       \cos(\Omega/2) & i\sin(\Omega/2)\\
       i\sin(\Omega/2) & \cos(\Omega/2)\\
    \end{array}
\right) \left(
    \begin{array}{c}
     \hat{a}_1 \\
     \hat{a}_2
    \end{array}
\right) \,,
\label{eq:moderot}
\end{eqnarray}
in agreement with our previous result (\ref{eq:hbeam}) for the effective
Hamiltonian of the real beam splitter (\ref{eq:rot}).
Consequently, the beam splitter performs rotations in the three-dimensional
space spanned by the quantum Stokes parameters (\ref{eq:js}).
Such rotations are independent on the overall phase $\Lambda/2$
in the factorization (\ref{eq:u2}).
For rotations on the quantum Poincar\'e sphere we can thus restrict
the beam-splitter transformations to SU(2) matrices \cite{Cornwell}
with
\begin{equation}
\mathrm{det}\underline{B} = 1\,.
\end{equation}
Finally,
we arrive at the general evolution operator for the beam splitter with
matrix (\ref{eq:u2})
\begin{equation}
\hat{B} = \exp(-i\Phi\hat{L}_z)\, \exp(-i\Theta\hat{L}_y)\,
\exp(-i\Psi\hat{L}_z)\, \exp(-i\Lambda\hat{L}_t)\,.
\end{equation}

\subsection{Wave-particle dualism}

Now we possess the theoretical tools to predict what happens
when two beams of quantum light interfere at a beam splitter. In
classical optics, the light beams are characterized by their
spatial shapes, by their normalized spatial mode functions $u_1$
and $u_2$, and by their amplitudes $a_1$ and $a_2$. The outgoing
modes have the amplitudes
\begin{equation}
\label{eq:class} \left(
    \begin{array}{c}
     a_1'  \\
     a_2'
    \end{array}
\right) = \underline{B} \left(
    \begin{array}{c}
     a_1  \\
     a_2
    \end{array}
\right)\,.
\end{equation}
The amplitudes of the incident light modes may statistically
fluctuate if the beams are not perfectly coherent
\cite{BornWolf,MandelWolf}. In this case the outgoing modes
fluctuate accordingly, because the individual amplitudes are
transferred according to the relation (\ref{eq:class}). In
quantum optics, the observables such as the amplitudes
$\hat{a}_1$ and $\hat{a}_2$ or the photon numbers
$\hat{a}_1^\dagger\hat{a}_1$ and $\hat{a}_2^\dagger\hat{a}_2$ may
statistically fluctuate in repeated experiments, even if the
light has always been prepared in identical pure states
\cite{Leonhardt}. Such quantum fluctuations tend to be quite
subtle and hard to discriminate from classical noise in
experiments. The quantum-noise properties distinguish the various
quantum states of light \cite{Leonhardt}. The coherent states
\cite{Leonhardt} resemble classical light beams with well-defined
amplitudes, coherent light. A coherent state of light is
characterized by the Wigner function \cite{Leonhardt}
\begin{equation}
\label{eq:wc} W(q,p) =
\frac{1}{\pi}\,\exp\left(-(q-q_0)^2-(p-p_0)^2\right) =
\frac{1}{\pi}\,\exp\left(-2|\,\alpha-a_0|^2\right) \,,
\end{equation}
where $a_0=(q_0+ip_0)/\sqrt{2}$ denotes the classical amplitude
of the light beam and $\alpha$ abbreviates $(q+ip)/\sqrt{2}$. The
Wigner function (\ref{eq:wc}) describes the quantum-statistical
fluctuations of the amplitude components $q$ and $p$, the
quadratures \cite{Leonhardt}. The vacuum state belongs to the
class of coherent states as well \cite{Leonhardt} --- the vacuum
is the coherent state with zero average amplitude. Yet the field
amplitudes of the vacuum state still fluctuate \cite{Leonhardt}.
We see from the Wigner function (\ref{eq:wc}) that the amplitudes
of coherent states fluctuate precisely like the quantum vacuum
around their average values $a_0$. The coherent states are the
most classical-like states of light, corresponding to waves as
perfect as quantum mechanics allows. Note that the number of
photons fluctuates in a coherent state, because precision in
amplitude and precision in particle number are mutually
exclusive. The photons in a coherent state of amplitude $a_0$ are
as randomly distributed as raisins in a cake with $|\,a_0|^2$
apiece \cite{Leonhardt}. Technically \cite{Leonhardt}, the photons
follow a Poisson distribution around the average $|\,a_0|^2$.

Suppose that the two light beams incident on the beam splitter
are in coherent states with the amplitudes $a_1$ and $a_2$,
corresponding to the two-mode Wigner function
\begin{equation}
\label{eq:win} W(\alpha_1,\alpha_2) = \frac{1}{\pi^2}\,
\exp\left(-2|\,\alpha_1-a_1|^2-2|\,\alpha_2-a_2|^2\right) \,.
\end{equation}
To predict the quantum state of the outgoing modes, we apply the
transformation rule (\ref{eq:wignertrans}) to the Wigner function
(\ref{eq:win}) and utilize the unitarity (\ref{eq:unib}) of the
beam-splitter matrix. We obtain
\begin{equation}
\label{eq:wout} W'(\alpha_1,\alpha_2) = \frac{1}{\pi^2}\,
\exp\left(-2|\,\alpha_1-a_1'|^2-2|\,\alpha_2-a_2'|^2\right) \,.
\end{equation}
The outgoing modes are in coherent states with the amplitudes
classically transformed (\ref{eq:class}). The modes are
completely uncorrelated, because the Wigner function factorizes.
The coherent states thus interfere just like classical waves,
even down to the finest details of their quantum-statistical
properties. This result uniquely distinguishes coherent states
\cite{Aharonov} and it can be extended to any passive optical
network \cite{Reck,Torma1,Torma3}. Historically, the interference
property of coherent states has been deduced from a microscopic
model of the beam splitter \cite{Brunner} and has served as the
starting point for the quantum theory of such optical instruments
\cite{PaulSplitter}.

Now, suppose that one incident beam carries precisely $n$ photons
and that no light impinges on the back of the semi-transparent
mirror. The light beam with exactly $n$ photons is in the Fock
state \cite{Leonhardt}
\begin{equation}
\label{eq:fock} |\,n\,\rangle =
\frac{1}{\sqrt{n!}}\,\hat{a}^{\dagger n} |\,0\,\rangle \,,
\end{equation}
and the other incident mode is in the vacuum state
$|\,0\,\rangle$. Fock states are the eigenstates of the
photon-number operator $\hat{a}^\dagger\hat{a}$ and hence they
correspond to light with a perfectly well-defined number of
photons \cite{Leonhardt}. We calculate the quantum state of the
outgoing modes
\begin{equation}
\hat{B}^\dagger |\,n\,\rangle_1\, |\,0\,\rangle_2 =
\frac{1}{\sqrt{n!}}\,\hat{B}^\dagger\hat{a}^{\dagger n}
|\,0\,\rangle_1\, |\,0\,\rangle_2 =
\frac{1}{\sqrt{n!}}\,\hat{B}^\dagger\hat{a}^{\dagger n}\hat{B}\,
|\,0\,\rangle_1\, |\,0\,\rangle_2 \,.
\end{equation}
Here we have used the fact that the beam splitter transforms the
incident vacuum into the outgoing vacuum, {\it ex nihilo nihil},
as we easily see from the Wigner function (\ref{eq:wout}) with
zero initial amplitudes $a_1$ and $a_2$. Since
\begin{equation}
\hat{B}^\dagger \left(
    \begin{array}{c}
     \hat{a}_1  \\
     \hat{a}_2
    \end{array}
\right) \hat{B} = \underline{B}^{-1} \left(
    \begin{array}{c}
     \hat{a}_1  \\
     \hat{a}_2
    \end{array}
\right) = \left(
    \begin{array}{cc}
      B_{11}^* & B_{21}^* \\
      B_{12}^* & B_{22}^*
    \end{array}
\right)\left(
    \begin{array}{c}
     \hat{a}_1  \\
     \hat{a}_2
    \end{array}
\right)\,,
\end{equation}
we obtain according to the Binomial theorem and the definition
(\ref{eq:fock}) of the Fock states
\begin{eqnarray}
\hat{B}^\dagger |\,n\,\rangle_1\, |\,0\,\rangle_2 &=&
\frac{1}{\sqrt{n!}}\,\left(B_{11}\,\hat{a}_1^\dagger +
B_{21}\,\hat{a}_2^\dagger\right)^n
\,|\,0\,\rangle_1\,|\,0\,\rangle_2 \nonumber\\
&=& \sum_{k=0}^n
\sqrt{
\left(\begin{array}{c}n\\k\end{array}\right)
}
B_{11}^k\,B_{21}^{n-k} \,|\,k\,\rangle_1\,|\,n-k\,\rangle_2 \,,
\label{eq:focksplit}
\end{eqnarray}
\begin{equation}
\left(\begin{array}{c}n\\k\end{array}\right) =
\frac{n!}{(n-k)!k!} \,.
\end{equation}
The beam splitter does not
split the incident photons, of course, but rather the
semi-transparent mirror statistically distributes the photons
into the reflected and the transmitted beam. Suppose we count the
photons in each emerging mode \cite{Brendel}. Each individual run
of the experiment \cite{Brendel} is unpredictable, but averaged
over a large statistical ensemble we get the joint photon-number
distribution
\begin{equation}
\label{eq:binomial} p(n_1,n_2) =
\left(\begin{array}{c}n\\k\end{array}\right)
\tau^{2k}\,(1-\tau^2)^{n-k}\,
\delta_{n_1,\,k}\,\delta_{n_2,\,n-k}\,,
\end{equation}
where $\tau$ denotes the transmissivity $|B_{11}|$. The Binomial
distribution (\ref{eq:binomial}) describes a random decision
process where $n$ distinguishable objects are distributed to two
channels, to the first channel with probability $\tau^2$ per
object and to the second one with probability $1-\tau^2$,
accordingly, because the beam splitter is assumed to be perfectly
lossless. Each photon is statistically independent, and so the
probability for $k$ individual photons to arrive in the first
channel and $n-k$ photons in the second one is the product
$\tau^{2k}\,(1-\tau^2)^{n-k}$. We multiply this value by the
Binomial coefficient, which describes the number of possibilities
to distribute {\it any} $k$ of the $n$ photons to the first
channel and the rest to the second one, because we cannot
discriminate between individual photons in
photon-counting experiments.
Nevertheless, photons behave in beam-splitting experiments as if
they were in-principle distinguishable,
in contrast to the common
statement that photons are fundamentally indistinguishable particles,
which illustrates some of the conceptional subtleties
of the photon.

There is another twist in the physics of photons and the beam
splitter. Suppose you let two beams of light with equal
intensities interfere at a perfect 50:50 beam splitter
characterized by the real matrix
\begin{equation}
\label{eq:5050} \underline{B} = \left(
    \begin{array}{cc}
     {1}/{\sqrt{2}} & {1}/{\sqrt{2}} \\
     -{1}/{\sqrt{2}}& {1}/{\sqrt{2}}
    \end{array}
\right) \,.
\end{equation}
Consider two coherent states with equal complex amplitudes
$a_0=a_1=a_2$. We obtain from the transformation rule
(\ref{eq:class}) that the first outgoing mode is in the coherent
state with complex amplitude $\sqrt{2}\, a_0$, whereas the second
mode is in the vacuum state. The two incident light beams
interfere constructively in the first outgoing mode and
destructively in the second one. If the two incident coherent
states have equal amplitudes but opposite phases, $a_1=a_0$,
$a_2=-a_0$, they interfere the other way round. Now, suppose you
let one single photon interfere with another single photon. We
calculate the quantum state of the outgoing modes,
\begin{eqnarray}
\hat{B}^\dagger |\,1\,\rangle_1\, |\,1\,\rangle_2 &=&
\frac{1}{2}\, \left(\hat{a}_1^\dagger-\hat{a}_2^\dagger\right)\,
\left(\hat{a}_1^\dagger+\hat{a}_2^\dagger\right)\,
|\,0\,\rangle_1\,|\,0\,\rangle_2
\nonumber\\
&=& \frac{1}{\sqrt{2}} \Big( |\,2\,\rangle_1\,|\,0\,\rangle_2 -
|\,0\,\rangle_1\,|\,2\,\rangle_2 \Big)\,. \label{eq:interfere}
\end{eqnarray}
The photons interfere constructively or destructively.
Complete destructive interference implies that the affected
outgoing mode is in the vacuum state, whereas the other mode must
carry exactly two photons, because the total number of photons is
conserved. Destructive or constructive interference depends on
the relative phase of the incident photons. Fock states with
precisely defined photon number are the most extreme particle-like
states of light, and hence they do not carry any wave-like phase
information. Faced with this dilemma, the beam splitter
distributes the two photons in either way with
$1/\sqrt{2}$ probability amplitude, {\it i.e.}\ with
$50\%$ probability after detection,
a truly Solomonic solution. Experimentally
\cite{Hong}, the coincident counts of photons reach a
well-pronounced minimum when the spatial-temporal modes of the
incident photons overlap at the beam splitter. The choice which
one of the outgoing mode carries the two photons becomes only
apparent when the light is detected. The same is true for the beam
splitting of $n$ photons discussed previously. The beam splitter
itself is a deterministic device. The probabilistic outcome of the
photocounting remains undecided until the measurement is made.
The quantum state (\ref{eq:interfere}) of the outgoing modes is
strongly correlated, and so is the state (\ref{eq:focksplit}) of
the split Fock state, in contrast to the interference
(\ref{eq:wout}) of coherent states. Moreover, the decisive
measurement devices may be located a long distance apart from
each other. Both photon-interference and photon-splitting
experiments are suitable \cite{Tan,Hardy} to test the
non-locality of quantum mechanics \cite{Bell,EPR}.

\section{Absorber and amplifier}

The optical instruments studied so far are completely reversible
devices. For example, when a Fock state, carrying a precise
number of photons, is split at a beam splitter we could, in
principle, send the outgoing beams back to restore the initial
Fock state. Mathematically argued, the instruments
(\ref{eq:linear}) are reversible, because the quasi-unitary
$\underline{S}$ matrix has the inverse
$\underline{G}\,\underline{S}^\dagger \underline{G}$, according
to Eq.\ (\ref{eq:qu}). So far, we have excluded irreversible
processes such as the absorption or the amplification of light.
However, all quantum processes are fundamentally
reversible, as long as no measurements are made or could be made
in principle (whatever measurement processes are), and as long as
we keep track of the quantum systems involved.

Consider for example an absorber, a piece of grey material. Some
of the incident light is destined for absorption and some part is
transmitted, with reduced intensity though. The absorbed component
is transferred to the material and ebbs away in many small
material excitations as heat. Normally we are simply not able to
keep track of the material details and so the absorbed quanta are
lost. (Exceptions are very simple atomic systems, two-level
systems for example, where absorption is reversible
\cite{JC,Paul0,ShoreKnight}.) The first part of the absorption process
resembles beam splitting. We could model the second, the
irreversible part by discarding the quantum information carried
in one of the outgoing beams, {\it i.e.}\ by averaging over the
unobserved component of the total quantum system. Modeling
absorption and detection losses by fictitious beam splitters has
been a successful idea in quantum optics
\cite{Fearn,JIL,Knoelletal,LPs,PaulReview,YuenShapiro}.
This theoretical trick is known elsewhere
as the thermo-field technique \cite{Umezawa}. In fact, many
absorbers are first of all scatterers, but it is quite remarkable
that we can sum up the multitude of scattered light modes in just
one outgoing mode of a fictitious beam splitter. Moreover, an
absorber may emit thermal radiation according to its temperature,
and so we should include emission as well as absorption in our
model. When the stimulated emission dominates the device acts as
an amplifier. To understand how to model such irreversible
processes requires some theory
\cite{BP,Carmichael,Gardiner,GardinerZoller,Lindblad}.

\subsection{Lindblad's theorem}

Lindblad \cite{Lindblad} determined the most general structure of
the dynamic equation for the density operator $\hat{\rho}$,
assuming only that the evolving $\hat{\rho}$ represents indeed an
ensemble of pure quantum states $|\,\psi_a\rangle$ occuring with
probabilities $p_a$. A reversible quantum process would only
change the pure states $|\,\psi_a\rangle$ while an irreversible
process may affect both the $|\,\psi_a\rangle$ and their
probabilities $p_a$. Lindblad's master equation, the quantum
version of the Boltzmann equation,
reads \cite{BP,Carmichael,Gardiner,GardinerZoller,Lindblad}
\begin{equation}
\label{eq:master}
\partial_t \hat{\rho} = \frac{i}{\hbar}\,[\hat{\rho},\hat{H}_0] -
\sum_l \gamma_l \left( \hat{L}_l^\dagger\hat{L}_l\hat{\rho} -
2\hat{L}_l\hat{\rho}\hat{L}_l^\dagger +
\hat{\rho}\hat{L}_l^\dagger\hat{L}_l \right) \,.
\end{equation}
Here $\hat{H}_0$ denotes the Hamiltonian of the reversible part
of the dynamics, while the $\gamma_l$ quantify the rates of the
irreversible processes described by the Lindblad operators
$\hat{L}_l$. The parameter $t$
may describe a fictitious time that, for example,
corresponds to the
penetration depth of an absorbing material or to the length
of a laser amplifier.
We define the effective Hamiltonian
\begin{equation}
\hat{H}_{\rm eff} = \hat{H}_0 -i\hbar \sum_l \gamma_l
\hat{L}_l^\dagger\hat{L}_l \,,
\end{equation}
a non-Hermitian operator, and write the master equation
(\ref{eq:master}) as
\begin{equation}
\label{eq:master2}
\partial_t \hat{\rho} =
\frac{i}{\hbar}\left(\hat{\rho}\hat{H}_{\rm eff}^\dagger -
\hat{H}_{\rm eff}\hat{\rho}\right) +
2\sum_l\gamma_l\hat{L}_l\hat{\rho}\hat{L}_l^\dagger  \,.
\end{equation}
The effective Hamiltonian alone would reduce the total quantum
probability $\mathrm{tr}\hat{\rho}$. The component
$i(\hat{\rho}\hat{H}_{\rm eff}^\dagger - \hat{H}_{\rm
eff}\hat{\rho})/\hbar$ of the master equation (\ref{eq:master2})
describes the coherent part of the irreversible process, for
example damping or amplification, while the terms
$2\gamma_l\hat{L}_l\hat{\rho}\hat{L}_l^\dagger$ characterize the
effect of quantum jumps \cite{BP,Carmichael}, fluctuations that
restore the nature of the density operator. In this respect,
Lindblad's theorem \cite{Lindblad} formulates the quantum version
of the fluctuation-dissipation theorem \cite{LL5}.

The Lindblad operators $\hat{L}_l$ describe the specific physical
effects of the irreversible processes involved in the dynamics
(\ref{eq:master}), the quantum transitions caused.
In short, the $\hat{L}_l$ are the transition operators.
For specific irreversible processes we can frequently use
our intuition to infer the relevant transition operators.
We may guess that the absorption of light corresponds
to the effect of the annihilation operator $\hat{a}$,
while the light emission
is represented by the creation operator $\hat{a}^\dagger$.
An absorber or amplifier is thus modeled by the
Lindblad operators
\begin{equation}
\label{eq:lindblad} \hat{L}_1 = \hat{a} \,,\quad  \hat{L}_2 =
\hat{a}^\dagger \,.
\end{equation}
For simplicity, we ignore the Hamiltonian $\hat{H}_0$ of the
single light mode that would only generate a time-dependent phase
shift of light. We translate the master equation
(\ref{eq:master}) with the Lindblad operators (\ref{eq:lindblad})
into the evolution equation of the Wigner function, a
Fokker-Planck equation
\cite{Carmichael,Gardiner,GardinerZoller,Risken}. For this, we
express the master equation in terms of the quadratures $\hat{q}$
and $\hat{p}$ with $\hat{a}=(\hat{q}+i\hat{p})/\sqrt{2}$ and
calculate the Wigner transforms (\ref{eq:wigner}) of the
operators involved. We use the correspondence rules
\cite{Gardiner}
\begin{eqnarray}
\hat{q}\hat{F}\longleftrightarrow
\left(q+\frac{i}{2}\,\partial_p\right)W_F&,&\quad
\hat{F}\hat{q}\longleftrightarrow
\left(q-\frac{i}{2}\,\partial_p\right)W_F\,, \nonumber\\
\hat{p}\hat{F}\longleftrightarrow
\left(p-\frac{i}{2}\,\partial_q\right)W_F&,&\quad
\hat{F}\hat{p}\longleftrightarrow
\left(p+\frac{i}{2}\,\partial_q\right)W_F\,,
\end{eqnarray}
between the operators and their Wigner transforms, and arrive at
the Fokker-Planck equation
\begin{equation}
\label{eq:fokker} \partial_t W = (\gamma_1-\gamma_2)
\big(\partial_q(qW)+\partial_p(pW)\big) +
\frac{\gamma_1+\gamma_2}{2} \big(\partial_q^2W+\partial_p^2W\big)
\,.
\end{equation}
The first term, with prefactor $\gamma_1-\gamma_2$, describes the
drift of the quasiprobabilities, to zero if the absorption
dominates and to infinity if the emission is stronger. The second
term, with rate $(\gamma_1+\gamma_2)/2$, describes the diffusion
of the quasiprobabilities due to quantum noise.

\subsection{Absorber}

Suppose that the absorption rate $\gamma_1$ outweighs the
emission rate $\gamma_2$. In this case, the Fokker-Planck
equation (\ref{eq:fokker}) describes the net effect of an
absorber. We find the stationary solution, normalized to unity,
\begin{equation}
\label{eq:wth} W_\mathrm{th}(q,p) = \frac{1}{\pi(2N+1)}\,
\exp\left(-\frac{q^2+p^2}{2N+1}\right) \,,\quad 2N+1 =
\left|\frac{\gamma_1+\gamma_2}{\gamma_1-\gamma_2}\right| \,.
\end{equation}
The Wigner function $W_\mathrm{th}$
corresponds to the thermal state
\cite{Leonhardt}
\begin{equation}
\label{eq:thermal}
\hat{\rho} = \frac{N}{N+1}\, \exp\left(-\frac{\hbar\omega\hat{a}^\dagger\hat{a}}
{k_\mathrm{B}T}\right)
\end{equation}
with average photon number $N$ and temperature
$T$, according to Planck's formula
\begin{equation}
\label{eq:planck} N =
\left[\exp\left(\frac{\hbar\omega}{k_\mathrm{B}T}\right) -
1\right]^{-1} \,.
\end{equation}
The stationary solution indicates that the absorber consists of a
thermal reservoir with temperature $T$, a reservoir that absorbs
light, but that also emits thermal radiation.
Now, consider the general solution of the  Fokker-Planck
equation (\ref{eq:fokker}). One verifies easily that the
solution is  \cite{LeoSU2}
\begin{equation}
\label{eq:fpsol} W(\alpha,t) =
\int_{-\infty}^{+\infty}\int_{-\infty}^{+\infty}
W(\alpha_1',t_0)\,W_\mathrm{th}(\alpha_2')\,
\mathrm{d}q_0\,\mathrm{d}p_0 \,,
\end{equation}
where the $\alpha$'s abbreviate the $(q+ip)/\sqrt{2}$ amplitudes,
with
\begin{equation}
\label{eq:absorber} \left(
    \begin{array}{c}
     \alpha_1'  \\
     \alpha_2'
    \end{array}
\right) = \left(
    \begin{array}{cc}
      \sqrt{\eta} & -\sqrt{1-\eta} \\
      \sqrt{1-\eta} & \sqrt{\eta}
    \end{array}
\right) \left(
    \begin{array}{c}
     \alpha  \\
     \alpha_0
    \end{array}
\right)
\end{equation}
and
\begin{equation}
\label{eq:eta} \eta = \exp\left[-2(\gamma_1-\gamma_2)(t-t_0)
\right] \,,\quad t \ge t_0 \,.
\end{equation}
Any initial Wigner function $W(\alpha,t_0)$ is exponentially
attenuated and eventually approaches the thermal state
$W_\mathrm{th}$ in the limit $t\rightarrow+\infty$.
The result (\ref{eq:fpsol}) with the relation (\ref{eq:absorber})
proves that partial absorption corresponds
to a simple beam splitter model. The incident light appears to be
split into the transmitted and the absorbed component. The $\eta$
parameter (\ref{eq:eta}) describes the transmission probability
of a single photon and $1-\eta$ is the probability of absorption.
The second mode of the fictitious beam splitter plays a double
role. The mode represents the reservoir into which the absorbed
light disappears and over which we average in the solution
(\ref{eq:fpsol}). Additionally, the initial state of the mode
describes the fluctuations of the thermal reservoir that
contaminate the transmitted light. In the Heisenberg picture we
represent the mode operator of the partially absorbed light as
\begin{equation}
\hat{a}(t) = \hat{a}(t_0) \sqrt{\eta} + \hat{a}_0 \sqrt{1-\eta}
\,.
\end{equation}
The fluctuation mode is essential in order to preserve the Bose
commutation relation of $\hat{a}(t)$, even at zero temperature.
Usually, for light in the optical range of the spectrum, room
temperature and zero temperature makes little difference. In this
case, the vacuum fluctuations of the reservoir affect the
partially absorbed light. If the light has initially
been in a coherent state with amplitude $a_0$ the transmitted
light remains in a coherent state with the reduced amplitude
$\sqrt{\eta}\,a_0$, despite the vacuum fluctuations,
because, as we know, the beam splitter
transforms coherent states (the initial state and the vacuum
mode) into disentangled coherent states. On the other hand, other
quantum states approach coherent states during the absorption
process. The absorber purifies light with excess amplitude noise,
but the absorber also destroys fragile non-classical states with
unusual quantum properties, such as Schr\"odinger-cat states
\cite{Brune,LeoSU2}.

\subsection{Amplifier}

Suppose that the emission rate $\gamma_2$ outweighs the
absorption rate $\gamma_1$ in the irreversible process
(\ref{eq:master}) with the Lindblad operators
(\ref{eq:lindblad}). In this case the Fokker-Planck equation
(\ref{eq:fokker}) has the general solution (\ref{eq:fpsol}) with
\cite{LeoSU11}
\begin{equation}
\label{eq:amp} \left(
    \begin{array}{c}
     \alpha_1'  \\
     \alpha_2'^*
    \end{array}
\right) = \left(
    \begin{array}{cc}
      \sqrt{\eta} & -\sqrt{\eta-1} \\
      -\sqrt{\eta-1} & \sqrt{\eta}
    \end{array}
\right) \left(
    \begin{array}{c}
     \alpha  \\
     \alpha_0^*
    \end{array}
\right)
\end{equation}
and with the $\eta$ parameter (\ref{eq:eta}) larger than unity.
The amplitude $\alpha$ of the initial Wigner function grows with
$\sqrt{\eta}$ and exponentially in $t$, which indicates that the
process describes a linear amplifier with gain
$\eta=\exp[2(\gamma_2-\gamma_1)(t-t_0)]$. The amplification is
accompanied by amplification noise, usually spontaneous-emission
noise in laser amplifiers, summed up in the thermal Wigner
function (\ref{eq:wth}) that becomes interwoven with the initial
quantum state. The noise temperature (\ref{eq:planck})
characterizes the quality of the amplifier. Amplification noise
is stronger than absorption noise in the sense that amplified
coherent states do not remain coherent states, even at zero noise
temperature.
Therefore, amplification does not simply reverse attenuation.
The growth of the signal mode combined with the inevitable
amplification noise appears in the Heisenberg picture as
\begin{equation}
\hat{a}(t) = \hat{a}(t_0) \sqrt{\eta} + \hat{a}_0^\dagger
\sqrt{\eta-1} \,.
\end{equation}
As in the case of the absorber, the fluctuation mode preserves the
Bose commutation relation of the amplified light, but the
amplification noise always creates additional quanta, indicated
by the creation operator, in contrast to the absorption noise.

Suppose that the amplifier attempts to balance the effect of
attenuation, {\it i.e.}\ $\gamma_1=\gamma_2=\gamma$ in the process
(\ref{eq:master}) with the Lindblad operators
(\ref{eq:lindblad}). Imagine, for example, that Eve, the eavesdropper,
tries to tap quantum information by beam splitting while covering
up her tracks by amplification. In the case when the emission
rate is equal to the absorption rate the Fokker-Planck equation
(\ref{eq:fokker}) reduces to the pure diffusion of the Wigner
function, without drift, as designed. We calculate the
quantum-statistical purity \cite{Gardiner,Leonhardt} of the signal
state using the overlap formula (\ref{eq:overlap})
\begin{equation}
\mathrm{tr}\{\hat{\rho}^2\} = 2\pi \int_{-\infty}^{+\infty}
\int_{-\infty}^{+\infty} W^2\, \mathrm{d}q\,\mathrm{d}p \,,
\end{equation}
and obtain from the Fokker-Planck equation (\ref{eq:fokker}) by
partial integration
\begin{equation}
\partial_t \mathrm{tr}\{\hat{\rho}^2\} = -4\pi\gamma
\int_{-\infty}^{+\infty}\int_{-\infty}^{+\infty}
\Big((\partial_qW)^2+(\partial_pW)^2\Big)
\mathrm{d}q\,\mathrm{d}p
< 0 \,.
\end{equation}
The purity $\mathrm{tr}\{\hat{\rho}^2\}$ monotonously decreases
until the Wigner function has been completely leveled by
diffusion, containing no information anymore. Eavesdropping
spoils the purity of the quantum state.

To summarize,
both the attenuation and the amplification of light correspond to
simple analog models that exactly describe the quantum effects of
such irreversible processes. An absorber is represented by a beam
splitter and an amplifier by a parametric amplifier. The second
mode of the beam splitter represents the absorption reservoir,
while the additional fluctuation mode of the amplifier describes
the amplification noise. We have proven \cite{LeoSU2,LeoSU11} the
equivalence between these simple models and the master equation
(\ref{eq:master}) for the Lindblad operators (\ref{eq:lindblad}),
{\it i.e.}\ for thermal reservoirs. We can easily extend
\cite{LeoSU2,LeoSU11} our analog models to phase-sensitive
Gaussian reservoirs \cite{Gardiner} with the
squeezed Lindblad operators
\begin{equation}
\label{eq:sqlindblad} \hat{L}_1 = \mu\hat{a}+\nu\hat{a}^\dagger
\,,\quad \hat{L}_2 = \hat{L}_1^\dagger \,,\quad |\mu|^2-|\nu|^2=1
\,,
\end{equation}
characterized by the complex constants $\mu$ and $\nu$,
see Refs.\ \cite{LeoSU2,LeoSU11} and Refs.\ cited therein.
The transformation (\ref{eq:sqlindblad}) squeezes the thermal Wigner
function (\ref{eq:wth}) of the quantum noise in one phase-space
direction and stretches it in the orthogonal direction,
indicating that the reservoir is indeed phase sensitive, possibly
with reduced fluctuations in one of the quadratures. The
fluctuation mode is in a state with Gaussian Wigner function and
Gaussian density operator \cite{Gardiner}. Whether our simple
models can be extended beyond Gaussian reservoirs remains unknown.

\section{Parametric amplifier}

The prime example of an active linear device is the optical
parametric amplifier \cite{Shen}. Phase-conjugating mirrors
\cite{Shen} and four-wave mixers \cite{Shen} belong to the same category.
The quantum optics of parametric amplifiers is studied in Refs.\
\cite{Caves,Huttner2,JIL,LeyLoudon,MG1,MG2,PaulFS,Townsend,YMK},
the quantum properties of phase-conjugating mirrors are considered in Refs.\
\cite{Agarwal1,Agarwal2,BAH,Gaeta,OBM} and the quantum
effects of four-wave mixers are studied in Refs.\
\cite{Kumar,ReidWalls,YS,Yurke}.
Active devices require external energy sources, often
provided by other light beams. These pump beams interact with the
modes to be amplified in non-linear media \cite{Bloembergen,Shen},
mostly certain crystals in the case of parametric amplifiers. The modes
are linearly amplified, as long as they do not feed back to the
pump processes. Here we focus entirely on the regime of linear
amplification. The  quantum physics of pump depletion has been
analyzed in Refs.\ \cite{Bandilla1,Bandilla2}.

The simplest example for parametric amplification in physics is
the playground swing. Rocking the legs changes the moment of
intertia. Rocking with twice the fundamental frequency of the
swing amplifies the oscillation, starting from tiny initial
movements, a phenomenon called parametric resonance \cite{LL1}.
The simplest optical example of a parametric amplifier is the
downconverter \cite{MandelWolf}.
Pump light with frequency $\omega_p$ drives two
other beams of light, called the signal and the idler, with
frequencies $\omega_s+\omega_i=\omega_p$. Assisted by the
non-linear medium, some pump photons with energy $\hbar\omega_p$
decay into photon pairs with energies $\hbar\omega_s$ and
$\hbar\omega_i$. The Hamiltonian (\ref{eq:hamp}) describes such a
process, the creation of photon pairs. The Hamiltonian also
accounts for the reverse process where photon pairs become
annihilated with their energies transferred back to the pump. The
relative phase between the signal, idler and pump beams decides
the direction of the process. Furthermore, momentum conservation
requires that the wave vectors ${\bf k}_p$ of the pump light
should equal the sum of the wave vectors of signal and idler,
${\bf k}_s$ and ${\bf k}_i$, a condition called phase matching
\cite{Shen}. Parametric amplifiers with high quantum-noise quality
and efficiency tend to take pure crystals, good resonators and a
number of ingenious experimental tricks.

\subsection{Matrix structure and squeezing}

The Hamiltonian (\ref{eq:hamp}) of the downconverter generates
the linear mode transformation (\ref{eq:linear}) with the
$\underline{S}$ matrix (\ref{eq:lorentz}). The mode operator of
the signal, say $\hat{a}_1$, is mixed with the Hermitian
conjugate of the idler, $\hat{a}_2^\dagger$. Simultaneously, the
idler operator $\hat{a}_2$ is mixed with the conjugate of the
signal, $\hat{a}_1^\dagger$. We would expect the same for
phase-conjugating mirrors \cite{Shen}. Let us assume the mode
transformation
\begin{equation}
\label{eq:amptrans} \left(
    \begin{array}{c}
     \hat{a}_1'  \\
     \hat{a}_2'^\dagger
    \end{array}
\right) = \underline{B} \left(
    \begin{array}{c}
     \hat{a}_1  \\
     \hat{a}_2^\dagger
    \end{array}
\right)\,,\quad \underline{B} = \left(
    \begin{array}{cc}
      B_{11} & B_{12} \\
      B_{21} & B_{22}
    \end{array}
\right)\,,
\end{equation}
corresponding to the $\underline{S}$ matrix
\begin{equation}
\label{eq:amps}
\underline{S} =
    \left(
    \begin{array}{cccc}
      B_{11}&0&0&B_{12}\\ 0&B_{22}^*&B_{21}^*&0\\
      0&B_{12}^*&B_{11}^*&0\\ B_{21}&0&0&B_{22}
    \end{array}
\right) \,.
\end{equation}
The mode transformation (\ref{eq:amptrans}) describes pure
amplification, without scattering. We require that the mode
operators of both the incident and the amplified light are proper
Bose operators, which results in the quasi-unitarity relation
(\ref{eq:qu}) of the $\underline{S}$ matrix. Explicitly, we
obtain
\begin{equation}
|\,B_{11}\,|^2-|\,B_{12}\,|^2=1\,,\quad
|\,B_{21}\,|^2-|\,B_{22}\,|^2=1\,,\quad
B_{11}^*B_{21}-B_{12}^*B_{22}=0\,.
\end{equation}
The general solution of these equations is
\begin{equation}
\underline{B} = e^{i\Lambda/2}\left(
    \begin{array}{cc}
      \cosh(\Theta/2)\,e^{i(\Psi+\Phi)/2} &
      \sinh(\Theta/2)\,e^{i(\Psi-\Phi)/2}  \\
      \sinh(\Theta/2)\,e^{i(\Phi-\Psi)/2}  &
      \cosh(\Theta/2)\,e^{-i(\Psi+\Phi)/2}
    \end{array}
\right)
\end{equation}
with the real parameters $\Phi$, $\Theta$, $\Psi$ and $\Lambda$.
We express $\underline{B}$ as the product
\begin{equation}
\label{eq:u11} \underline{B} = e^{i\Lambda/2} \left(
    \begin{array}{cc}
      e^{i\Psi/2} & 0 \\
      0 & e^{-i\Psi/2}
    \end{array}\right) \left(
    \begin{array}{cc}
      \cosh(\Theta/2) & \sinh(\Theta/2) \\
      \sinh(\Theta/2) & \cosh(\Theta/2)
    \end{array}
\right)\left(
    \begin{array}{cc}
      e^{i\Phi/2} & 0 \\
      0 & e^{-i\Phi/2}
    \end{array}\right)\,.
\end{equation}
Like in the case of the beam splitter, the net result of the
parametric amplifier or of the phase-conjugating mirror amounts
to four steps. First, the incident beams gain a phase $\Phi/2$,
yet in contrast to the beam splitter, this is not a relative
phase, but an absolute phase shift, because the transformation
(\ref{eq:amptrans}) acts on $\hat{a}_1$ and $\hat{a}_2^\dagger$,
not on $\hat{a}_1$ and $\hat{a}_2$. After the phase shift the
modes are amplified by the factor $\cosh(\Theta/2)$ and mixed,
with the overlap $\sinh(\Theta/2)$, and finally the outgoing modes
gain the absolute phase $\Psi/2$ and the relative phase
$\Lambda$. We could include the phases into the definitions of
the incident and the amplified modes. The hyperbolic mixing
(\ref{eq:lorentz}) would remain as the key feature of the device.
We can express the $\underline{S}$ matrix (\ref{eq:lorentz}) with
$\zeta=\Theta/2$ as
\begin{equation}
\label{eq:factors}
\underline{S} = \underline{R}^{-1}
    \left(
    \begin{array}{cccc}
      \cosh(\Theta/2)&0&\sinh(\Theta/2)&0\\
      0&\cosh(\Theta/2)&0&-\sinh(\Theta/2)\\
      \sinh(\Theta/2)&0&\cosh(\Theta/2)&0\\
      0&-\sinh(\Theta/2)&0&\cosh(\Theta/2)
    \end{array}
\right) \underline{R}
\end{equation}
in terms of the matrix of the 50:50 beam splitter (\ref{eq:5050})
\begin{equation}
\underline{R} = \left(
    \begin{array}{cccc}
     {1}/{\sqrt{2}}&{1}/{\sqrt{2}}&0&0\\
     -{1}/{\sqrt{2}}&{1}/{\sqrt{2}}&0&0\\
     0&0&{1}/{\sqrt{2}}&{1}/{\sqrt{2}}\\
     0&0&-{1}/{\sqrt{2}}&{1}/{\sqrt{2}}
    \end{array}
\right) \,.
\end{equation}
In the case when the signal and the idler modes are the linear
polarization modes of a single light wave the matrix
$\underline{R}$ describes the rotation of the polarization axis
by $\pi/4$. The parametric amplifier thus processes the rotated modes
separately
\begin{equation}
\hat{a}_{r1}' =
\hat{a}_{r1}\cosh(\Theta/2)+\hat{a}_{r1}^\dagger\sinh(\Theta/2)
\,,\quad \hat{a}_{r2}' =
\hat{a}_{r2}\cosh(\Theta/2)-\hat{a}_{r2}^\dagger\sinh(\Theta/2)
\,.
\end{equation}
We obtain for the quadratures (\ref{eq:quad})
\begin{equation}
\hat{q}_{r1}'=\hat{q}_{r1}\,e^{+\Theta/2}\,,\quad
\hat{p}_{r1}'=\hat{p}_{r1}\,e^{-\Theta/2}\,,\quad
\hat{q}_{r2}'=\hat{q}_{r2}\,e^{-\Theta/2}\,,\quad
\hat{p}_{r2}'=\hat{p}_{r2}\,e^{+\Theta/2}\,.
\label{eq:quadsqueeze}
\end{equation}
The $\hat{q}_{r1}$ quadrature is stretched and the $\hat{p}_{r1}$
quadrature is squeezed, while preserving the Heisenberg
commutation relation (\ref{eq:heisen}), as expected from a
canonical transformation. In the second rotated mode the
$\hat{q}_{r2}$ quadrature is squeezed and the $\hat{p}_{r2}$
quadrature is stretched accordingly. Therefore, in the rotated
basis, the parametric amplifier acts as a perfectly noiseless
amplifier or de-amplifier, a squeezer, for particular quadrature
components. The parametric amplifier may produce squeezed light
\cite{Breitenbach,Leonhardt,LoudonKnight}.
A reduction by up to $-7\mbox{dB}=20\%$
in the quadrature variance $\Delta^2 q$
compared with the vacuum noise has been
observed so far \cite{Lam}.

\subsection{Effective Lorentz transformations}

\begin{figure}
\begin{center}
\includegraphics[width=7cm]{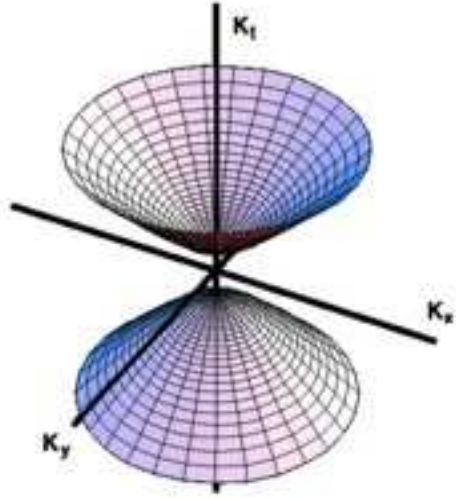}
\end{center}
\caption{\label{fig:hyper}
Hyperboloid of the parametric amplifier, the counterpart of the
Poincar\'e sphere of the beam splitter shown in Fig.\ \ref{fig:sphere}.
The picture schematically illustrates the $K$ operators
(\ref{eq:k}).
Depending on the sign of the photon-number difference
$\hat{K}_z+1/2$, the amplifier operates on the upper (positive)
or the lower (negative) sheet of the hyperboloid.
Amplification processes draw trajectories on one
of the sheets, consisting of the Lorentz transformations
(\ref{eq:lorentzmodes}).}
\end{figure}

The beam splitter (\ref{eq:bs}) generates abstract
three-dimensional rotations, expressed in terms of the Euler
angles $\Phi$, $\Theta$ and $\Psi$ in the decomposition
(\ref{eq:u2}) of the scattering matrix. To recall a less abstract
example, polarizers perform rotations in the Poincare sphere, on
the quantum Stokes parameters in the Jordan-Schwinger
representation (\ref{eq:js}). The parametric amplifier
(\ref{eq:amptrans}) turns out to generate effective Lorentz
transformations. We define the operators \cite{YMK}
\begin{eqnarray}
\hat{K}_t &=& \frac{1}{2}\, (\hat{a}_1^\dagger,\, \hat{a}_2)\,\,
\mathds{1}\, \left(
    \begin{array}{c}
     \hat{a}_1  \\
     \hat{a}_2^\dagger
    \end{array}
\right) = \frac{1}{2}\,\left(\hat{a}_1^\dagger\hat{a}_1 +
\hat{a}_2\hat{a}_2^\dagger\right)\,, \nonumber\\
\hat{K}_x &=& \frac{1}{2}\, (\hat{a}_1^\dagger,\, \hat{a}_2)\,
\sigma_x \left(
    \begin{array}{c}
     \hat{a}_1  \\
     \hat{a}_2^\dagger
    \end{array}
\right) = \frac{1}{2}\,\left(\hat{a}_1\hat{a}_2 +
\hat{a}_1^\dagger\hat{a}_2^\dagger\right)\,, \nonumber\\
\hat{K}_y &=& \frac{1}{2}\, (\hat{a}_1^\dagger,\, \hat{a}_2)\,
\sigma_y \left(
    \begin{array}{c}
     \hat{a}_1  \\
     \hat{a}_2^\dagger
    \end{array}
\right) = \frac{i}{2}\,\left(\hat{a}_1\hat{a}_2 -
\hat{a}_1^\dagger\hat{a}_2^\dagger\right)\,, \nonumber\\
\hat{K}_z &=& \frac{1}{2}\, (\hat{a}_1^\dagger,\, \hat{a}_2)\,
\sigma_z \left(
    \begin{array}{c}
     \hat{a}_1  \\
     \hat{a}_2^\dagger
    \end{array}
\right) = \frac{1}{2}\,\left(\hat{a}_1^\dagger\hat{a}_1 -
\hat{a}_2\hat{a}_2^\dagger\right)\,,
\label{eq:k}
\end{eqnarray}
with our notation deviating slightly from Ref.\ \cite{YMK}. We
see from the Bose commutation relation that
$\hat{K}_t=\hat{L}_t+1/2$ and $\hat{K}_z=\hat{L}_z-1/2$ where the
$L$ operators belong to the Jordan-Schwinger representation
(\ref{eq:js}). The $K$ operators (\ref{eq:k}) obey the
commutation relations
\begin{equation}
[\hat{K}_x,\,\hat{K}_y]=-i\hat{K}_t \,,\quad
[\hat{K}_y,\,\hat{K}_t]=i\hat{K}_x \,,\quad
[\hat{K}_t,\,\hat{K}_x]=i\hat{K}_y \,,
\end{equation}
while $\hat{K}_z$ commutes with all others. We find
\begin{equation}
\label{eq:properk} \hat{K}_t^2 - \hat{K}_x^2 - \hat{K}_y^2 =
\hat{K}_z(\hat{K}_z + 1) \,,
\end{equation}
the equivalent of the squared angular momentum (\ref{eq:totall}).
The $K$ operators lie on a hyperboloid, see Fig.\ \ref{fig:hyper}.
Since $\hat{K}_z$ commutes with $\hat{K}_t$, $\hat{K}_x$ and
$\hat{K}_y$, the $K$ operators generate transformations with the
invariant (\ref{eq:properk}), the quantum analog of the squared
space-time distance \cite{LL2} in 2+1 dimensions. In other words,
the $K$ operators generate effective Lorentz transformations. In
fact, we find
\begin{eqnarray}
\exp(-i\Theta\hat{K}_y) \left(
    \begin{array}{c}
     \hat{K}_t \\
     \hat{K}_x \\
     \hat{K}_y
    \end{array}
\right) \exp(i\Theta\hat{K}_y) &=& \left(
    \begin{array}{ccc}
      \cosh\Theta & \sinh\Theta & 0\\
      \sinh\Theta & \cosh\Theta & 0\\
      0 & 0 & 1
    \end{array}
\right) \left(
    \begin{array}{c}
     \hat{K}_t \\
     \hat{K}_x \\
     \hat{K}_y
    \end{array}
\right)\,, \nonumber\\
\exp(-i\Omega\hat{K}_x) \left(
    \begin{array}{c}
     \hat{K}_t \\
     \hat{K}_x \\
     \hat{K}_y
    \end{array}
\right) \exp(i\Omega\hat{K}_x) &=& \left(
    \begin{array}{ccc}
      \cosh\Omega & 0 & -\sinh\Omega\\
      0 & 1 & 0\\
      -\sinh\Omega & 0 & \cosh\Omega
    \end{array}
\right) \left(
    \begin{array}{c}
     \hat{K}_t \\
     \hat{K}_x \\
     \hat{K}_y
    \end{array}
\right)\,, \nonumber\\
\exp(-i\Phi\hat{K}_t) \left(
    \begin{array}{c}
     \hat{K}_t \\
     \hat{K}_x \\
     \hat{K}_y
    \end{array}
\right) \exp(i\Phi\hat{K}_t) &=& \left(
    \begin{array}{ccc}
      1 & 0 & 0\\
      0 & \cos\Phi & \sin\Phi\\
      0 & -\sin\Phi & \cos\Phi
    \end{array}
\right) \left(
    \begin{array}{c}
     \hat{K}_t \\
     \hat{K}_x \\
     \hat{K}_y
    \end{array}
\right)\,,
\end{eqnarray}
two Lorentz transformations \cite{LL2} with effective velocities
$v_x/c=\tanh\Theta$ and $v_y/c=\tanh\Omega$
and one rotation with angle $\Phi$.
We obtain for the mode operators
\begin{eqnarray}
\exp(-i\Theta\hat{K}_y) \left(
    \begin{array}{c}
     \hat{a}_1 \\
     \hat{a}_2^\dagger
    \end{array}
\right) \exp(i\Theta\hat{K}_y) &=& \left(
    \begin{array}{cc}
      \cosh(\Theta/2) & \sinh(\Theta/2)\\
      \sinh(\Theta/2) & \cosh(\Theta/2)\\
    \end{array}
\right) \left(
    \begin{array}{c}
     \hat{a}_1 \\
     \hat{a}_2^\dagger
    \end{array}
\right) \,, \nonumber\\
\exp(-i\Omega\hat{K}_x) \left(
    \begin{array}{c}
     \hat{a}_1 \\
     \hat{a}_2^\dagger
    \end{array}
\right) \exp(i\Omega\hat{K}_x) &=& \left(
    \begin{array}{cc}
       \cosh(\Omega/2) & i\sinh(\Omega/2)\\
       -i\sinh(\Omega/2) & \cosh(\Omega/2)\\
    \end{array}
\right) \left(
    \begin{array}{c}
     \hat{a}_1 \\
     \hat{a}_2^\dagger
    \end{array}
\right) \,, \nonumber\\
\exp(-i\Phi\hat{K}_t) \left(
    \begin{array}{c}
     \hat{a}_1 \\
     \hat{a}_2^\dagger
    \end{array}
\right) \exp(i\Phi\hat{K}_t) &=& \left(
    \begin{array}{cc}
      e^{i\Phi/2} & 0\\
      0 & e^{-i\Phi/2}\\
    \end{array}
\right) \left(
    \begin{array}{c}
     \hat{a}_1 \\
     \hat{a}_2^\dagger
    \end{array}
\right) \,,
\label{eq:lorentzmodes}
\end{eqnarray}
These are all transformations that belong to the class (\ref{eq:amptrans}).
Therefore, parametric amplifiers generate effective Lorentz
transformations. The measured quadrature-noise reduction
\cite{Lam} of $7\mbox{dB}$ ($20\%$)
gives, according to Eq.\ (\ref{eq:quadsqueeze}),
a $\Theta$ parameter of $-\log 0.2$, which corresponds to
an effective velocity of $v/c=\tanh \Theta=0.92$.
Finally, we employ the Lorentz generators (\ref{eq:k})
to represent the evolution
operator of the parametric amplifier with the matrix (\ref{eq:u11}).
We use our results (\ref{eq:lorentzmodes}) and get
\begin{equation}
\label{eq:ampevolve}
\hat{B} = \exp(-i\Phi\hat{K}_t)\, \exp(-i\Theta\hat{K}_y)\,
\exp(-i\Psi\hat{K}_t)\, \exp(-i\Lambda\hat{K}_z)\,.
\end{equation}
Passive optical instruments such as beam splitters preserve the
total number of photons (\ref{eq:totalnumber}), they
conserve energy. The parametric amplifier is subject to a
conservation law as well. The generators of the evolution
operator (\ref{eq:ampevolve}) commute with $\hat{K}_z$.
Consequently, the photon-number difference is conserved,
\begin{equation}
\label{eq:diff}
\hat{a}_1'^\dagger\hat{a}_1' - \hat{a}_2'^\dagger\hat{a}_2' =
\hat{a}_1^\dagger\hat{a}_1 - \hat{a}_2^\dagger\hat{a}_2 \,.
\end{equation}
Photons are emitted in pairs, one in the signal and one in the
idler beam, as we would expect from the Hamiltonian (\ref{eq:hamp}).

\subsection{Quantum correlations}

Conservation laws manifest themselves in correlations. For example,
picture a pair of two particles with spin, say two polarized photons.
The particles are produced at one spot and then they move away from
each other. Suppose that the total spin of the two particles has been zero
initially and that spin is conserved. Now, if we measure the spin of
the first particle, the second one must have the opposite spin,
regardless which type of polarization we are probing, linear or circular,
and regardless how far apart the particles are. Such long-ranging
correlations exist due to conservation laws, they do not violate the
relativistic causality, because they do not cause each other, but rather
have a cause in common, and they are completely classical.
Quantum mechanics adds a subtlety, a decisive yet quantitatively
subtle feature. For quantum particles the outcomes of the spin
measurements may be probabilistic, depending on their state,
but the spins of the partners are correlated. For example,
consider the singlet state of two photons in two light beams,
with two orthogonal polarization modes $(+)$ and $(-)$ each,
\begin{equation}
\label{eq:singlet0}
|\psi\rangle
= \frac{1}{\sqrt{2}}
\big(|+\rangle_{1}|-\rangle_{2}-
\big(|-\rangle_{1}|+\rangle_{2}\big) \,,
\end{equation}
or, written as Fock states,
\begin{equation}
\label{eq:singlet}
|\psi\rangle
= \frac{1}{\sqrt{2}}
\big(|1\rangle_{1+}|1\rangle_{2-}-
\big(|1\rangle_{1-}|1\rangle_{2+}\big) \,.
\end{equation}
The polarization of photon 1 is completely random, $(+)$ or $(-)$
with 50\% chance, but the polarization of the second
photon is always correlated to the first one.
The quantum state (\ref{eq:singlet0}) is entangled --- the vector
$|\psi\rangle$ of the total state
does not factorize into the state vectors of the subsystems.
Now, suppose we measure the polarization of the first
photon, with respect to a given axis, and we turn the polarizer
of the second by an angle. In this case, the measurement
results are not perfectly correlated anymore.
However, we can quantify the correlation degree
by calculating the difference $D$ between the
statistical frequency of coincidences and the frequency of
detecting the opposite spin,
after many repeated runs of the experiment
for each setting of the polarizer angle.
Any spin measurement is characterized by the scalar
product of a three-dimensional unity vector ${\bf a}$ with
the vector of the Pauli matrices
${\bf \sigma} = (\sigma_x, \sigma_y, \sigma_z)^T$
of Eq.\ (\ref{eq:pauli}), where
${\bf a}$ describes the polarizer setting on the Poincare sphere.
We obtain for the singlet state (\ref{eq:singlet0})
\begin{equation}
\label{eq:corr}
D({\bf a}_1, {\bf a}_2)=
\langle\psi\,|\, ({\bf a}_1 \cdot {\bf \sigma}_1)\,
({\bf a}_2 \cdot {\bf \sigma}_2)\,|\,\psi\rangle
= -{\bf a}_1 \cdot {\bf a}_2 \,.
\end{equation}
It turns out \cite{Bellpaper,Bell}
that $D$ violates certain inequalities for local hidden variables,
based on classical statistics,
called Bell's inequalities after their discoverer \cite{Bellpaper}.
Suppose that the particles are classical by nature, but are engaged
in a quantum conspiracy.
Before they are separated they agree on the outcome of
spin measurements denoted in
some hypothetical hidden variables that determine the
measurement results.
Although the observer is unaware of the plot,
he can put constraints on the correlator,
for example \cite{CHHS,CS}
\begin{equation}
\label{eq:chhs}
D({\bf a}_1, {\bf a}_2) +
D({\bf a}_1, {\bf a}_2') +
D({\bf a}_1', {\bf a}_2) -
D({\bf a}_1', {\bf a}_2') \le 2\,,
\end{equation}
assuming that the particles
cannot communicate with each other after they have been separated.
The spin correlations (\ref{eq:corr}) of the quantum state
(\ref{eq:singlet0}) violate the Bell inequality (\ref{eq:chhs})
by maximally a factor of $\sqrt{2}$, as one may verify.
Quantum particles can be slightly stronger correlated than
classical statistics allows.
Violations of Bell's inequalities have been convincingly
demonstrated in several experiments
since the major breakthrough in efficiency due to Aspect
\cite{Aspect}, see e.g. Refs.\ \cite{Tittel,Weihs}.
The parametric downconverter \cite{MandelWolf,Shen}
is the central tool in most
modern tests of genuine quantum correlations.
Apart from probing the foundations of quantum mechanics,
quantum correlations are beginning to play a decisive role in
tasks beyond the capabilities of classical physics
such as quantum cryptography \cite{Bruss,Gisin}
and in other forms of quantum communication
\cite{BEZ,NielsenChuang}.
Parametric amplifiers are frequently applied in this wider field
of entanglement engineering.

The debate on quantum correlations, on the
``spooky action at a distance'', that preceded any applications
of entanglement, dates back to the 1935 paper \cite{EPR}
``Can Quantum-Mechanical Description of Physical Reality
Be Considered Complete'' by Einstein, Podolsky and Rosen.
Schr\"odinger \cite{Schroedinger}
introduced the term entanglement to physics,
{\it Verschr\"ankung}, a German term used by
cabinet makers for dovetailing, quite a fitting term for the
quantum correlations in the natural world.
The paper \cite{EPR} by Einstein, Podolsky and Rosen
has spawned a large literature \cite{Mermin}
and has inspired a series of intriguing experiments,
see for example Refs.\
\cite{Aspect,Boschi,Bouwmeester,BEZ,Ou1,Ou2,Tittel,Weihs}.
Interestingly, the parametric amplifier represents not only
the device of choice for many modern applications of
quantum correlations, but the amplifier is also able
to reproduce the original argument
by Einstein, Podolsky and Rosen
for the quadratures $q$ and $p$
\cite{Reid} in the experiment \cite{Ou1,Ou2}.

Suppose the signal and idler modes are in the vacuum state
initially. An optical parametric amplifier is just an
amplifier after all. So, if the initial amplitude is zero the
amplified amplitude remains zero. However, amplitude
fluctuations get amplified. Remember the classic example of
parametric resonance \cite{LL1}, the
child on the playground swing where rocking at twice
the frequency of the swing amplifies tiny initial movements.
In the case of the optical parametric amplifier,
the vacuum fluctuations are amplified.
The resulting quantum state, called the two-mode squeezed vacuum,
comes close to the original state considered by Einstein, Podolsky
and Rosen in their 1935 paper \cite{EPR}.
We calculate its Wigner function.
Initially, the two modes are in the vacuum state with the
Wigner function \cite{Leonhardt}
\begin{equation}
W(q_1,p_1,q_2,p_2) = \frac{1}{\pi^2}\,
\exp\left(-q_1^2-p_1^2-q_2^2-p_2^2\right) \,.
\end{equation}
We recall that simple optical instruments transform the Wigner function
as if $W$ were a classical probability distribution of the quadratures.
For simplicity, we assume that the $\underline{S}$ matrix
(\ref{eq:amps}) of the amplifier is the real matrix (\ref{eq:lorentz}).
We obtain from the transformation rule (\ref{eq:wignertrans})
the Wigner function of the two-mode squeezed vacuum
\begin{equation}
W'(q_1,p_1,q_2,p_2) =
\frac{1}{\pi^2}\, \exp\left(-q_1'^2-p_1'^2-q_2'^2-p_2'^2\right)
\end{equation}
with
\begin{eqnarray}
\left(
    \begin{array}{c}
     q_1'  \\
     q_2'
    \end{array}
\right)& =&
\left(
    \begin{array}{cc}
      \cosh\zeta & \sinh\zeta \\
      \sinh\zeta & \cosh\zeta
    \end{array}
\right)
\left(
    \begin{array}{c}
     q_1  \\
     q_2
    \end{array}
\right)\,,
\nonumber\\
\left(
    \begin{array}{c}
     p_1'  \\
     p_2'
    \end{array}
\right) &=&
\left(
    \begin{array}{cc}
      \cosh\zeta & -\sinh\zeta \\
      -\sinh\zeta & \cosh\zeta
    \end{array}
\right)
\left(
    \begin{array}{c}
     p_1  \\
     p_2
    \end{array}
\right)\,.
\end{eqnarray}
Consequently,
\begin{equation}
\label{eq:wtmsv}
W' =
\frac{1}{\pi^2}\, \exp\left[
-\frac{(q_1-q_2)^2}{2e^{-2\zeta}}
-\frac{(q_1+q_2)^2}{2e^{2\zeta}}
-\frac{(p_1+p_2)^2}{2e^{-2\zeta}}
-\frac{(p_1-p_2)^2}{2e^{2\zeta}}
\right] \,.
\end{equation}
The Wigner function (\ref{eq:wtmsv}) indicates strong
quadrature correlations when $|\zeta|$ is large.
Suppose we measure the $q$ quadrature
of the first beam using homodyne detection \cite{Leonhardt}
and we find the value $q_1$. In this case a quadrature
measurement of the second beam, some distance away from
the first one, gives with high probability $q_1$ as well.
The first measurement has prepared the second
beam in a state that closely resembles a $q$ quadrature eigenstate.
Now, if we measure the $p$ quadratures, with the result $p_1$
for the first beam, the second mode produces $-p_1$ with
high probability.
The $q$ and $p$ eigenstates cannot coexist, because of the
canonical commutation relations (\ref{eq:heisen}).
Therefore, depending on the measurements
on the first beam, the second one is prepared in states
that are mutually incompatible
and that cannot coexist as ``elements of reality'' \cite{EPR}.
Of course, Einstein, Podolsky and Rosen
did not mention quantum-optical quadratures in their paper \cite{EPR},
but rather position and momentum in real space,
yet this does not fundamentally alter the argument.
However, a hypothetical quantum conspiracy could
explain the behavior
Einstein, Podolsky and Rosen found so puzzling.
The light beams could have originally agreed on with which
probability they are going to produce all linear combinations of
$q$ and $p$ quadratures. The Wigner function (\ref{eq:wtmsv})
reveals the plot. The function plays the role of the non-negative
probability distribution for some hidden variables triggering
all quadrature detection results. In our case, the quadratures
themselves are the hidden variables
$q_1$, $q_2$, $p_1$ and $p_2$.
Ironically, the paradox that led to quantitative measures of
quantum nonlocality can be explained classically.
It would take a negative Wigner function \cite{LeoVac}
to violate Bell's inequalities in quadrature space,
as Bell has pointed out \cite{BellEPW}, although with
flawed mathematics in his counterexample \cite{Johansen}.
However, the two-mode squeezed vacuum state
may well violate Bell's inequalities for other observables
than the quadratures \cite{Banaszek},
because the quasiprobability distributions
in other variables \cite{LeoDiscrete0,LeoDiscrete,Vacarro}
may be negative.

The parametric amplifier generates strong quantum correlations
between the emerging light beams. In fact, one can argue
\cite{BP1,BP2} that the correlations are the strongest possible
for two equal oscillator modes with given average energy.
To understand why, we calculate the reduced Wigner function
of any of the modes, say mode 1.
We obtain after Gaussian integration
\begin{equation}
\int_{-\infty}^{+\infty}\int_{-\infty}^{+\infty}
W'(q_1,p_1,q_2,p_2) \,\mathrm{d}q_2\,\mathrm{d}p_2 =
W_\mathrm{th}(q_1,p_1) \,,
\end{equation}
where $W_\mathrm{th}$ denotes the Wigner function
(\ref{eq:wth}) of the thermal state (\ref{eq:thermal})
with the temperature
\begin{equation}
\label{eq:tzeta}
T = \frac{\hbar\omega}{2k_\mathrm{B}\ln\coth\zeta} \,.
\end{equation}
The reduced Wigner function describes the reduced quantum state
of one of the correlated modes, averaged over the second mode.
As we know from statistical physics \cite{LL5}, the thermal state
is the state of maximal entropy with given energy,
in the canonical Gibbs ensemble, the state of maximal disorder
and minimal information.
On the other hand, the total two-mode squeezed vacuum is a pure
state. Consequently, this state stretches mostly across
the correlations between the two entangled modes.
When reduced to one mode, the subsystem contains as little
information as possible. Therefore, the two-mode squeezed
vacuum is indeed the most strongly correlated quantum states
of two equal harmonic oscillators with given energy \cite{BP1,BP2}.
The higher this energy the higher is the effective temperature
(\ref{eq:tzeta}) and the stronger are the quantum correlations.

Since the parametric amplifier conserves the photon-number
difference (\ref{eq:diff}) between signal and idler,
the quantum correlations of the two-mode squeezed vacuum
are particularly transparent in terms of photon-number states,
{\it i.e.}\ in the Fock representation.
We utilize the disentangling theorem \cite{Gilles,Perelomov}
\begin{equation}
\exp(2i\zeta\hat{K}_y) =
\exp(\hat{a}_1^\dagger\hat{a}_2^\dagger\tanh\zeta)\,
\exp(-2\hat{K}_t\ln\cosh\zeta)\,
\exp(-\hat{a}_1\hat{a}_2\tanh\zeta)\,,
\end{equation}
and get for the two-mode squeezed vacuum state
\begin{eqnarray}
|\psi\rangle =
\exp(2i\zeta\hat{K}_y)\, |0\rangle_1 |0\rangle_2
&=&
\exp(\hat{a}_1^\dagger\hat{a}_2^\dagger\tanh\zeta)\,
\exp(-2\hat{K}_t\ln\cosh\zeta)\,
|0\rangle_1 |0\rangle_2
\nonumber\\
&=&
\frac{1}{\cosh\zeta} \sum_{n=0}^\infty \frac{1}{n!}\,
(\hat{a}_1^\dagger\hat{a}_2^\dagger\tanh\zeta)^n\,
|0\rangle_1 |0\rangle_2
\nonumber\\
&=&
\frac{1}{\cosh\zeta} \sum_{n=0}^\infty
(\tanh\zeta)^n\,
|n\rangle_1 |n\rangle_2 \,.
\label{eq:tmsv}
\end{eqnarray}
We see that $|\psi\rangle$ consists entirely of photon pairs
$|n\rangle_1 |n\rangle_2$.
We also see that the reduced density operator
$\mathrm{tr}_2\{|\psi\rangle\langle\psi|\}$ is indeed
the thermal state (\ref{eq:thermal}) with temperature
(\ref{eq:tzeta}).
In spontaneous parametric downconversion \cite{MandelWolf}
the squeezing parameter $\zeta$ is small.
In this regime, $\zeta^2$ gives the probability for producing
a single pair of photons,
as we see from a first-order perturbation theory with
the Hamiltonian (\ref{eq:hamp}),
and $\zeta^{2n}$ gives the probability for $n$
independently produced pairs.
So, according to the result  (\ref{eq:tmsv}), the two-mode
squeezed vacuum seems to consist of
indistinguishable photon pairs,
in the regime of weak amplification.
Spontaneous downconversion \cite{MandelWolf} has been
frequently used for generating single pairs of photons,
because the probability for multiple pairs is very low,  $\zeta^{2n}$
compared with  $\zeta^2$.
High-efficiency single-photodetectors operate like Geiger counters,
where a single photon triggers an electron-hole avalanche
\cite{Leonhardt}. Since the initial photons are just triggers,
the detectors
cannot discriminate between one or more detected photons
in each detection click.
Therefore, it is an experimental advantage to guarantee that
mostly single pairs occur, which of course limits the rate of pair
production.

The interference experiments with single photons
mentioned in Sec.\ 4.3 have been performed with photon pairs
generated in spontaneous parametric downconversion
\cite{MandelWolf}.
Here the quantum state (\ref{eq:tmsv}) of light is essentially
\begin{equation}
|\psi\rangle \approx
|0\rangle_1 |0\rangle_2
+\zeta\,|1\rangle_1 |1\rangle_2 \,.
\end{equation}
In such experiments only those experimental runs count where
photons are counted, the time when the detectors are not firing
is ignored, which reduces the quantum state to the photon pair
$|1\rangle_1 |1\rangle_2$. Postselection disentangles the
two-mode squeezed vacuum. We argued in Sec.\ 4.3
that the interference of the photon pair
$|1\rangle_1 |1\rangle_2$ at a 50:50 beam splitter
generates the entangled state (\ref{eq:interfere}).
Without postselection, however, this state is the disentangled
product of two single-mode squeezed vacua, as we see from
the factorization (\ref{eq:factors}) of the $\underline{S}$ matrix.
The notion of entanglement is to some extent relative.

Finally, we explain briefly how the singlet state (\ref{eq:singlet})
is generated in spontaneous parametric downconversion
\cite{MandelWolf}. The spin singlet (\ref{eq:singlet}) refers
to the polarization state of two light beams with two polarization
modes each, four modes in total with the mode operators
$\hat{a}_{1+}$, $\hat{a}_{1-}$, $\hat{a}_{2+}$ and
$\hat{a}_{2-}$. The singlet is completely unpolarized, but
contains polarization correlations, Bell correlations \cite{Bell}
in fact. Therefore, the creation process of the singlet
must be polarization invariant. The Hamiltonian of parametric
downconversion is quadratic in the mode operators
of signal and idler, as long as the pump can be regarded
as a classical parameter. Consequently, we need a quadratic form
of the annihilation operators $\hat{a}_{1+}$, $\hat{a}_{1-}$,
$\hat{a}_{2+}$ and $\hat{a}_{2-}$ that is invariant under
polarization transformations \cite{Lehner}.
The rotations on the Poincare sphere (\ref{eq:poincarerot})
correspond to the mode transformations (\ref{eq:moderot})
of the beam-splitter type (\ref{eq:bs}) of the polarization modes
$(\hat{a}_{1+}, \hat{a}_{1-})^T$ and
$(\hat{a}_{2+}, \hat{a}_{2-})^T$, with identical SU(2) matrices
\cite{Cornwell}. Written in matrix form
\begin{equation}
\underline{\hat{a}}'=\underline{B}\,\underline{\hat{a}}
\,,\quad
\underline{\hat{a}} =
\left(
    \begin{array}{cc}
      \hat{a}_{1+} & \hat{a}_{2+} \\
      \hat{a}_{1-} & \hat{a}_{2-}
    \end{array}
\right)
\,,\quad
\underline{B}^{-1}=\underline{B}^\dagger
\,,\quad
\mathrm{det}\underline{B}=1\,.
\end{equation}
The determinant of the $\underline{\hat{a}}$ matrix has the desired
property, because $\mathrm{det}\underline{\hat{a}}'=
\mathrm{det}\underline{B}\,\mathrm{det}\underline{\hat{a}}$.
Consequently, we arrive at the effective Hamiltonian
\begin{equation}
\label{eq:hh}
\hat{H} =
i e^{i\varphi}\zeta\,\mathrm{det}\underline{\hat{a}}^\dagger -
i e^{-i\varphi}\zeta\,\mathrm{det}\underline{\hat{a}}
\end{equation}
with real squeezing parameter $\zeta$ and real phase $\varphi$.
We assume that both signal and idler have initially been in the
vacuum state. We use the result (\ref{eq:tmsv}) and get
\begin{eqnarray}
|\psi\rangle &=& \exp(-i\hat{H})\,
|0\rangle_{1+}|0\rangle_{1-}|0\rangle_{2+}|0\rangle_{2+}
= |\psi\rangle_+ + |\psi\rangle_- \,,
\nonumber\\
|\psi\rangle_\pm &=&
\frac{1}{\cosh\zeta} \sum_{n=0}^\infty
\left(\pm e^{-i\varphi}\tanh\zeta\right)^n\,
|n\rangle_{1\pm} |n\rangle_{2\mp} \,,
\label{eq:polinv}
\end{eqnarray}
which, after postselecting for single photon pairs, gives
the Bell state (\ref{eq:singlet}) in the regime of spontaneous
parametric downconversion.
We describe briefly how the
downconverter can realize the Hamiltonian (\ref{eq:hh}).

\begin{figure}
\begin{center}
\includegraphics[width=10cm]{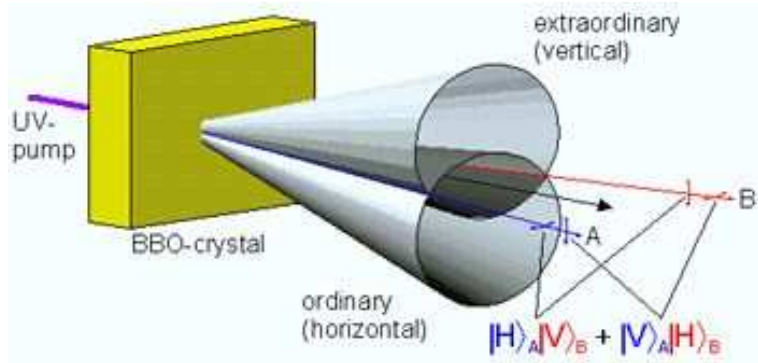}
\end{center}
\caption{\label{fig:double}
Schematic of the downconversion process
(with type-II phase matching).
An ultraviolet photon incident on a nonlinear crystal
can sometimes spontaneously split into
two correlated infrared photons.
These photons are emitted on opposite sides of the pump beam,
along two cones,
one of which has horizontal polarization,
the other of which has vertical polarization.
Photon pairs emitted along the intersections
of the cones are entangled in polarization ---
each photon is individually unpolarized,
and yet the photons necessarily have
perpendicular polarizations,
no matter how far apart they are.
(Courtesy of Paul Kwiat.)}
\end{figure}
\begin{figure}
\begin{center}
\includegraphics[width=6cm]{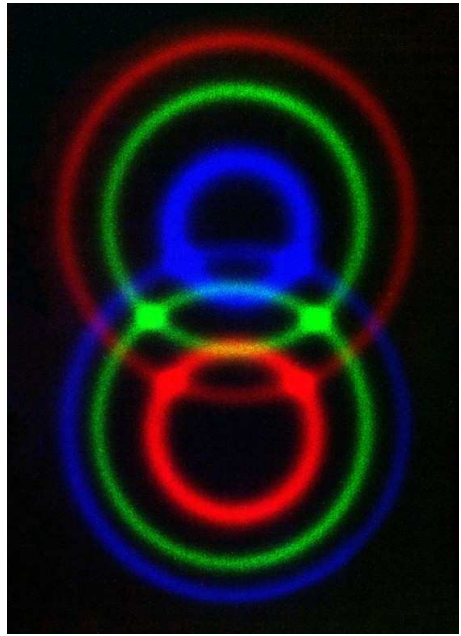}
\end{center}
\caption{\label{fig:cones}Entangled light. Colorized infrared
photograph of the measured downconversion output \cite{Kwiat1}
schematically shown in Fig. \ref{fig:double}. The green rings
correspond to photons with roughly equal energies (half that of
the parent pump photon). Where the rings intersect the
polarization state of the light is entangled. (Courtesy of Anton
Zeilinger, Institut f\"ur Experimentalphysik, Universit\"at Wien.)}
\end{figure}

Downconversion is a nonlinear optical process
\cite{Bloembergen,Shen}
where a pump beam $E_p$ of frequency
$\omega_p$ generates two waves, the signal  $E_s$ and
the idler $E_i$, of frequencies
$\omega_s$ and $\omega_i$ with $\omega_p=\omega_s+\omega_i$.
This three-wave mixing
\cite{Shen} requires an appropriately anisotropic medium, a crystal,
because the interaction Hamiltonian of the three fields
$\chi^{(2)} E_p E_s E_i$ is not invariant under parity transformations
where the fields change sign.
In an anisotropic medium light propagates as ordinary or as
extraordinary waves with opposite polarizations \cite{BornWolf}.
There are two principal types of downconverters.
In type I \cite{Shen} the modes $E_s$ and $E_i$
are both ordinary or extraordinary waves.
In type II downconversion \cite{Shen} either $E_s$ or $E_i$
is ordinary and the other wave is extraordinary, and hence the two
have opposite polarizations.
The downconverted light leaves the crystal in two cones
that display the conservation laws involved ---
the conservation of energy, $\omega_p=\omega_s+\omega_i$,
and the conservation of the wave vectors
(phase matching condition \cite{Shen}),
see Fig.\ \ref{fig:double}.
Where the cones intersect the polarization state is undecided.
According to our theory, this polarization-invariant state of light
is the state (\ref{eq:polinv}). The two intersection lines of the
emerging light cones carry polarization Bell states \cite{Kwiat1},
see Fig.\ \ref{fig:cones}.
Alternatively, one can employ two subsequent type I crystals
with orthogonal optical axes \cite{Kwiat2}.
The crystals downconvert pump light into
entangled beams in all directions.
All this shows that the parametric amplifier is a superb device
for generating quantum correlations of light.

\section{Optical black hole}

\begin{figure}
\begin{center}
\includegraphics[width=12cm]{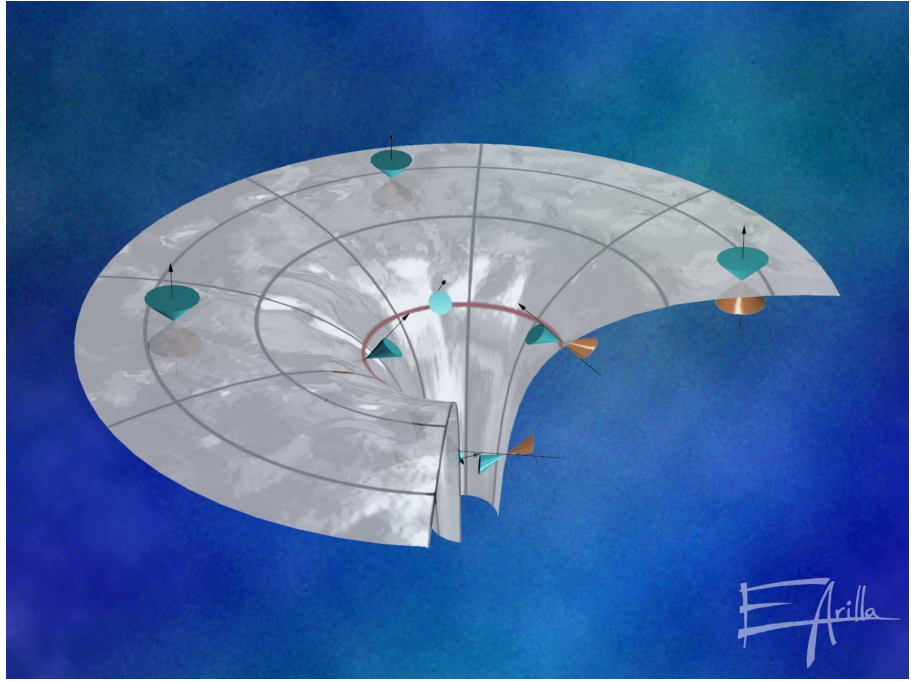}
\end{center}
\caption{\label{fig:cc}Black hole. Gravity modifies the metric of
space and time. In flat space, far away from the gravitating
object, the light cones stand upright in a space-time diagram.
Gravity is tilting the light cones, bending light. At the event
horizon all future light cones point
inwards. Light cannot escape anymore, and so cannot anything else.
(Picture by Enrique Arilla.) }
\end{figure}

Black holes are optical instruments as well, although they tend to operate
on grander scales than ordinary laboratory equipment.
First of all, black holes are gravitational lenses \cite{Schneider} ---
the gravitational fields surrounding them focus light from other sources
and may cause a multitude of optical illusions.
Black holes share this feature with other gravitating bodies such as
entire galaxies, stars or even individual planets.
The distinctive feature of a black hole is the event horizon \cite{MTW}.
The horizon, surrounding the hole, is a place of no return.
Beyond the horizon, everything falls into the hole,
no matter how fast it moves, even light.
Here gravity has tilted the future light cones such that
they all point inwards, see Fig.\ \ref{fig:cc}.

As we discuss in this section, the horizon turns the black hole into an
active optical instrument. The hole does not merely act as a
gravitational lens, it emits radiation as well, Hawking radiation
\cite{Birrell,Brout,Hawking1,Hawking2}.
The hole appears as a thermal black body with a temperature that
depends on the gravity at the horizon (and hence \cite{MTW}
on the horizon's area and the hole's mass).
The characteristic wavelength at the peak of the Planck spectrum is
comparable to the radius of the horizon \cite{Birrell,Brout},
more than $1\mbox{km}$
for the known solar-mass or larger black holes,
which makes Hawking radiation far too cold to be observable
against the cosmic microwave background
of about $1\mbox{mm}$ wavelength.

On the other hand, since black holes are optical instruments after all,
why not use optical equipment to make one?
Such table-top black holes would be much smaller than
the astronomical ones. Since the scale of the hole determines the
temperature of the Hawking radiation, one would expect them to
radiate stronger than their larger archetypes in space.
Most of the proposed artificial black holes \cite{Novello}
are based on a simple idea \cite{Unruh,Visser}:
Consider a moving medium with spatially varying flow speed, say
water going down the drain of a bathtub.
Waves in the fluid, sound waves \cite{Unruh,Visser},
surface waves \cite{Schuetzhold2} or light waves
\cite{Brevik,LeoPiwstor,Schuetzhold1} are trapped
when the flow exceeds the speed of the wave in the medium,
the speed of sound or the  effective speed of light, for example.
To turn this simple analogy into measurements of emergent
Hawking radiation will take extraordinary media though ---
Bose-Einstein condensates of alkali vapours
\cite{Dalfovo,PS} for sonic holes \cite{Garay,LKO},
two phases of superfluid Helium-3 \cite{VolovikBook}
for ripple holes \cite{Schuetzhold2} and slow-light media
\cite{LeoSlow,Lukin}
for optical holes \cite{LeoPiwliten,LeoNature,LeoCatastrophe}.
Note that the possibly realistic version of the optical hole
\cite{LeoNature,LeoCatastrophe}
differs from ordinary black holes.
It does not involve a moving medium and belongs
to a different class of quantum catastrophes \cite{LeoCatastrophe}
than the Hawking effect
\cite{Birrell,Brout,Hawking1,Hawking2}.
Much in this field is still in flux and is connected to other areas
of physics outside of the scope of this article.
Here we discuss an optical {\it Gedankenexperiment},
not a realistic proposal, to show how Hawking radiation
emerges in principle.

\subsection{Light in moving media}

Consider a dielectric medium with an incredibly strong electric
permittivity $\varepsilon \gg 1$ and with unity magnetic permeability,
for simplicity. Such a medium would significantly reduce the
phase velocity of light, $c'$, because
\begin{equation}
c'^2 = \frac{c^2}{\varepsilon} \,.
\end{equation}
Note that the slow light \cite{LeoSlow,Lukin} demonstrated
in recent experiments \cite{Hau,Liu,Philips} represents
a different case where the group velocity of light is reduced
and not the phase velocity.
Consider for simplicity a one-dimensional medium moving
with the spatially varying flow speed $u(x)$.
The Lagrangian of the electromagnetic field
is the Lorentz scalar \cite{LL2,LeoSpace}
\begin{equation}
{\mathscr L}=
\frac{1}{2} \left(ED - BH\right) \,.
\end{equation}
At any place in the moving medium we can imagine a co-moving
frame of coordinates where the medium is at rest,
denoted by primes. The Lagrangian is invariant,
\begin{equation}
{\mathscr L}=
\frac{1}{2} \left(E'D' - B'H'\right)  \,.
\end{equation}
In a local co-moving frame, the $D'$ and $H'$ fields
are connected to the $E'$ and $B'$ fields via the
constitutive equations \cite{Jackson,LL8}
\begin{equation}
D' = \varepsilon_0\varepsilon E' ,\quad
H' = \varepsilon_0 c^2 B'  \,.
\end{equation}
We represent the $E'$ and $B'$ fields in terms (\ref{eq:vectorpotential})
of the vector potential $A$, a scalar in 1+1 dimensions.
For non-relativistic velocities of both the medium and the light
we obtain from the local Galilei transformations to the laboratory frame
\begin{equation}
\partial_t' = \partial_t + u \partial_x ,\quad \partial_x' = \partial_x \,.
\end{equation}
In this way we arrive at the Lagrangian in the laboratory frame
\begin{equation}
\label{eq:lmove}
{\mathscr L}=\frac{\varepsilon_0\varepsilon}{2}
\Big((\partial_t A)^2 + 2u(\partial_t A)(\partial_x A) +
(u^2-c'^2)(\partial_x A)^2\Big)
\end{equation}
that generates the Euler-Lagrange equation \cite{LL2}
\begin{equation}
\label{eq:wavemove}
\partial_t \varepsilon (\partial_t + u\partial_x) A +
\partial_x \varepsilon [ u\partial_t + (u^2 - c'^2)]\partial_x A = 0 \,.
\end{equation}
This wave equation describes the propagation of classical light
in moving media in one spatial dimension.

\subsection{Space-time geometry}

Remarkably, light turns out to experience the moving medium
as an emergent space-time geometry
\cite{Gordon,LeoPiwstor,LeoSpace,Nov,PhamMauQuan1,PhamMauQuan2}.
To understand why, consider the simplest case, our one-dimensional
medium with large $\varepsilon$.
We introduce the relativistic notation
\begin{equation}
x^0 = t,\quad x^1 = x,\quad
\partial_0=\partial_t,\quad\partial_1=\partial_x
\end{equation}
and adopt Einstein's summation convention over repeated indices.
In this way we write the Lagrangian (\ref{eq:lmove}) as
\begin{equation}
{\mathscr L}=\frac{\varepsilon_0}{2}\,
f^{\mu\nu} (\partial_\mu A)(\partial_\nu A)
\end{equation}
in terms of the matrix
\begin{equation}
  f^{\mu \nu} = \varepsilon \left(
    \begin{array}{cc}
      1 & u \\
      u & -c'^2  + u^2
    \end{array}
\right) \, ,
\end{equation}
which gives the Euler-Lagrange equation
\begin{equation}
\label{eq:wavef}
\partial_\mu f^{\mu\nu} \partial_\nu A = 0 \,.
\end{equation}
Suppose that $\varepsilon$ is constant while $u$ may vary in space.
We introduce the metric
\begin{equation}
\label{eq:metric}
  g_{\mu \nu} = \left(
    \begin{array}{cc}
      c'^2-u^2 & u \\
      u & -1
    \end{array}
\right)
\end{equation}
with the determinant \cite{LL2}
\begin{equation}
g = -c'^2
\end{equation}
and the inverse
\begin{equation}
  g^{\mu \nu} = \frac{1}{c'^2} \left(
    \begin{array}{cc}
      1 & u \\
      u & -c'^2  + u^2
    \end{array}
\right) \, .
\end{equation}
As long as $\varepsilon$ remains constant we can write the wave equation
(\ref{eq:wavef}) as \cite{LL2}
\begin{equation}
\label{eq:dalembert}
0 = \frac{1}{\sqrt{-g}}\, \partial_\mu \sqrt{-g}\, g^{\mu\nu} \partial_\nu A
= D_\mu D^\mu A \,.
\end{equation}
The $D_\mu$ denote the covariant derivatives and $D_\mu D^\mu$
gives the D'Alembert operator in general coordinates
with the metric (\ref{eq:metric}).
Equation (\ref{eq:dalembert}) describes a mass-less field, light,
in general relativity. Therefore,
the electromagnetic field perceives the moving medium as the
space-time geometry (\ref{eq:metric}).
Light-rays follow zero-geodesic lines measured with respect to the metric
(\ref{eq:metric}) \cite{LeoPiwstor}.
In the history of optics, the oldest known paradigm of geometric guidance
is Fermat's Principle \cite{BornWolf}:
Light rays follow the shortest optical paths where the path length
is measured with respect to the refractive index.

A dielectric medium acts on light, in focusing or scattering light,
but, according to the principle of
{\it actio et reactio}, light also acts on the medium,
in exerting optical forces.
We have ignored these forces here,
assuming that the medium is very heavy.
The light forces turn out to follow geometric ideas, too.
The medium perceives the electromagnetic field as a space-time
geometry as well. Light and dielectric matter see each other
as geometries \cite{LeoSpace}.

Finally, the idea that media modify the geometry of space and time
can be turned around, at least speculatively.
Can one interpret space-time curvature, {\it i.e.}\ gravity,
as generated by a hypothetical fluid, as an emergent phenomenon
in condensed-matter physics \cite{VolovikBook}?
Sir Isaac Newton speculated
in 1675 \cite{Brewster}
that gravity acts through ``an aetherial medium,
much of the same constitution with air, only the vibrations
far more swift and minute'', one of Sir Isaac's departures to
hypotheses. Much later, in 1968, Andrei Sakharov,
father of the Soviet hydrogen bomb, dissident and Nobel
Peace Price laureate, noticed \cite{Sacharov}
that Einstein's general relativity
could appear as the effect of fluctuations of the quantum
vacuum, the modern form of the aether.

\subsection{Horizon}

The effective metric of light in moving media \cite{Gordon,LeoSpace}
does not cover all possible space-time manifolds, but the
generated geometries are rich enough to include horizons.
The horizon is the place where the flow speed $u$ exceeds the
speed of light in the medium, $c'$.
Here, according to the metric (\ref{eq:metric}),
the measure of time, $c'^2-u^2$, is zero --- time appears to stand still.
Light propagating against the current freezes.
In fact, black holes have been termed ``frozen stars'' \cite{Harrison}.
Immediately before the horizon, any outgoing light struggles to escape.
The closer the distance to the horizon, the more optical cycles
it takes to move forward. In turn, the wave length of light
shrinks to zero at the horizon.
On the other side of the horizon, no counter-propagating light
is able to escape anymore.

Let us turn these words into formulae.
Assume for simplicity that $\varepsilon$ is constant.
In this case the general solution of the wave equation (\ref{eq:wavef}) is
\begin{equation}
\label{eq:axt}
A(x,t) = A_+(\tau_+) + A_-(\tau_-) \,,\quad
\tau_\pm = t - \int\frac{\mathrm{d}x}{u \pm c'} \,.
\end{equation}
Note that also the complex conjugate, $A^*$, is a solution,
because the wave equation (\ref{eq:wavef}) is real.
Suppose, without loss of generality, that the medium flows from the
right to the left such that $u$ is negative.
Hence $A_+(\tau_+)$ describes counter-propagating and $A_-(\tau_-)$
co-propagating wavepackets. The $\tau_\pm$ are the null coordinates
in the global medium frame. (Note that in more than one spatial
dimension such a frame does not exist in general.)
Assume that the velocity $u$ exceeds the effective speed of light at
one point, say $x=0$, such that
\begin{equation}
u \sim -c' + \alpha x \quad \mbox{near}\,\, x=0 \,.
\end{equation}
The positive constant $\alpha$ describes the velocity gradient
at the horizon, the analogue of the surface gravity
at the event horizon of a gravitational hole.
In the vicinity of the horizon we get
\begin{equation}
\tau_+ \sim t - \frac{\ln (x/x_{\pm\infty})}{\alpha}
\end{equation}
with the integration constants $x_{\pm\infty}$.
Therefore, to escape from the horizon
takes an exponentially long time.
Assuming monochromatic light, where
$A_+=\exp(-i\omega\tau_+)$,  we see that
the phase $\varphi$ of the light wave is logarithmic in $x/x_{\infty}$.
Consequently, the wavelength $2\pi/(\partial_x\varphi)$
is indeed proportional to the distance from the horizon,
\begin{equation}
\lambda \sim \lambda_{\infty} \frac{x}{x_{\infty}} \,.
\end{equation}
We also see that the $\tau_+$ coordinate distinguishes
the left and the right side of the horizon.
The constants $x_{\pm\infty}$ must be positive on the right
and negative on the left side, for not running into problems
with complex time.
The horizon cuts space into two disconnected parts.

\subsection{Hawking radiation}

Consider the quantum theory of light for our artificial black hole,
the moving medium.
As usual \cite{Birrell,Weinberg},
the quantum field $\hat{A}$ consists of a set of modes
with the mode functions $A_k$ and the Bose annihilation
operators $\hat{a}_k$, see Eq.\  (\ref{eq:modeex}).
In the moving medium, the mode functions are subject
to the wave equation (\ref{eq:wavef}).
The modes are normalized according to a scalar product
that ought to stay invariant in time and that should approach the
scalar product (\ref{eq:scalar}) in the limiting case when
the medium is at rest. We put
\begin{eqnarray}
\left(A_1,A_2\right) &\equiv&
\frac{i\varepsilon_0}{\hbar}
\int \left(A_1^*\,f^{0\nu}\partial_\nu\, A_2 -
A_2 \,f^{0\nu}\partial_\nu \,A_1^*\right)
\mathrm{d}x
\nonumber\\
&=&
\frac{i\varepsilon_0}{\hbar}
\int \Big(A_1^*\,(\partial_t + u\partial_x) A_2 -
A_2 \,(\partial_t + u\partial_x)A_1^*\Big) \,\varepsilon\,
\mathrm{d}x \,.
\label{eq:scalarmove}
\end{eqnarray}
The scalar product (\ref{eq:scalarmove}) is the desired one,
because it agrees with Eq.\  (\ref{eq:scalar}) when $u$ vanishes
and the product is indeed a constant in time,
because of the conservation law
\begin{equation}
\label{eq:consmove}
\partial_\mu \left(A_1^*\, f^{\mu\nu} \partial_\nu A_2 -
A_2\, f^{\mu\nu} \partial_\nu A_1^*\right) = 0 \,.
\end{equation}
After these general remarks on quantum light in moving media
we return to the optical black hole.

Consider the monochromatic modes with frequency $\omega$
that are propagating against the current, the mode function
$A_R$ on the right side of the horizon and
the function $A_L$ on the left side,
\begin{equation}
\label{eq:alar}
A_R = {\cal A}_R\, \Theta(x)\,\exp(-i\omega\tau_+) \,,\quad
A_L = {\cal A}_L\, \Theta(-x)\,\exp(i\omega\tau_+) \,.
\end{equation}
The unity step functions $\Theta(\pm x)$ indicate that
$A_R$ is only supported on the right side, whereas
$A_L$ exists solely on the left side.
Note that the mode $A_L$ on the left side beyond the horizon
oscillates with the negative frequency $-\omega$,
in order to represent a proper positive-norm mode
(\ref{eq:orthonorm}). In fact, we see from
\begin{equation}
i(\partial_t + u\partial_x) A_L =
-\omega\,\frac{c'}{u+c'}\, A_L
\end{equation}
that the normalization integral (\ref{eq:scalarmove})
is positive for negative $u+c'$, as is the case beyond the horizon.
Negative frequencies $\omega$ correspond to negative energies
$\hbar\omega$. Consequently, the artificial black hole
does not have a natural ground state, at least in our simplified
model where the permittivity $\varepsilon$ and the flow $u$
are given and fixed, ignoring the backaction of the light
onto the medium.
The vacuum state, the state of zero photons,
traditionally the ground state, may thus depend on the history
of the horizon, on the catastrophic event creating the black hole.
In space, the gravitational collapse creates
the black hole. In our laboratory analogue, the medium
must have been accelerated in the past to form a stationary flow
with a horizon. Consider the quantum theory of light in the
Heisenberg picture where the field operators evolve and the
quantum state of light remains constant.
Trace the mode functions back into their pre-horizon history,
in order to find the decisive signature of the modes
that distinguishes the initial quantum vacuum, and hence
the final vacuum, too.
Complex analysis provides us with
elegant theoretical tools for this purpose
\cite{Brout,DamourRuffini}.

Before the horizon has been formed, the vacuum modes stretched
over the entire medium. Counter-propagating waves consisted
of superpositions of plane waves with positive wavenumbers,
\begin{equation}
A_0(x,t)\sim \int_0^{+\infty} \widetilde{A}_0(k)\,
\exp(ikx - i\omega t)\,\mathrm{d}k \,,\quad
t \rightarrow -\infty \,.
\end{equation}
Consider the analytic continuation of the vacuum modes
to complex $x$ values. The $A_0$ are analytic on the upper
half plane, because here $\exp(ikx)$ decays for positive
wavenumbers $k$, such that $A_0$ cannot develop
singularities. Now, imagine that the original modes evolve during the
formation of the horizon where the medium is gently
accelerated to exceed the effective speed of light beyond one point.
A smooth process cannot fundamentally alter the analytic
properties of waves.
Consequently, we can employ the analyticity of the mode functions
as a marker to distinguish the vacuum state of light.
The modes $A_R$ and $A_L$ are clearly non-analytic, because
they lack support on either one of the horizon sides.
Yet we can glue $A_R$ and $A_L^*$ together to form
analytic wave functions \cite{Brout}.
Close to the horizon, we have
\begin{equation}
\exp(-i\omega\tau_+) \sim (x/x_{+\infty})^{i\omega/\alpha}
\exp(-i\omega t) \,.
\end{equation}
Circumventing the horizon on the upper half plane, we use the property
$(-z)^{i\mu} = e^{-\pi\mu} z^{i\mu}$,
and find the combinations
\begin{equation}
\label{eq:bogo}
A_1 = A_R \cosh\zeta + A_L^*\sinh\zeta \,,\quad
A_2 = A_R^* \sinh\zeta + A_L\cosh\zeta
\end{equation}
that are analytic on the upper half plane when
\begin{equation}
\tanh\zeta = e^{-\pi\omega/\alpha} \,.
\end{equation}
The vacuum modes $A_1$ and $A_2$ are normalized according
to Eq.\ (\ref{eq:orthonorm}) with the scalar product
(\ref{eq:scalarmove}), taking into account
that both $A_R$ and $A_L$ are normalized and orthogonal,
and that the norm of $A^*$ is the negative norm of $A$.
The mode transformations (\ref{eq:bogo}) imply that
\begin{equation}
\left(
    \begin{array}{c}
     \hat{a}_R  \\
     \hat{a}_L^\dagger
    \end{array}
\right) =
\left(
    \begin{array}{cc}
      \cosh\zeta & -\sinh\zeta \\
      -\sinh\zeta & \cosh\zeta
    \end{array}
\right)
\left(
    \begin{array}{c}
     \hat{a}_1  \\
     \hat{a}_2^\dagger
    \end{array}
\right) \,.
\end{equation}
The black hole acts as a parametric amplifier, transforming
the incident vacuum into the two-mode squeezed vacuum
state (\ref{eq:tmsv}), the entangled Einstein-Podolsky-Rosen
state \cite{EPR}. The steeper the velocity gradient $\alpha$
the stronger is the entanglement and hence the energy of the
two-mode squeezed vacuum.
The horizon generates photon pairs,
both propagating against the current,
but one escaping on the right side of the horizon
and the other drifting to the left.
Somehow the flow must provide the energy for the
photon production that, in turn, will cause friction.
In space \cite{MTW},
the gravitational field surrounding the black hole,
created by the mass of the collapsed object,
should provide the energy for the emerging Hawking radiation.
The black hole is evaporating and is getting smaller.
All black holes are similar, apart from their scales \cite{MTW}.
Therefore, smaller holes generate larger gradients of the gravitational
potential. Consequently \cite{Hawking1},
the evaporation is accelerating,
until the black hole explodes in a grand fiery finale.

If we confine ourselves to the right side of the horizon, our
natural place in the case of gravitational black holes,
we perceive the two-mode squeezed vacuum (\ref{eq:tmsv})
as the thermal state (\ref{eq:thermal}) with the temperature
(\ref{eq:tzeta}), the Hawking temperature
\begin{equation}
\label{eq:hawking}
T = \frac{\hbar\alpha}{2\pi k_\mathrm{B}} \,.
\end{equation}
The temperature depends only on the velocity gradient $\alpha$,
the analogue of the surface gravity of the hole.
Consequently, the Hawking temperature (\ref{eq:hawking})
is the same for the entire spectrum, which is highly
non-trivial. The black hole appears as a black-body
radiator in thermal equilibrium
\cite{Bekenstein1,Bekenstein2,Brout,JP},
a mysterious result \cite{Hawking1}.

\subsection{Grey-body factor}

We have shown that the black hole acts as an active optical
instruments, as a parametric amplifier.
So far, we have ignored the passive optical properties of the hole,
the analogue of the gravitational lens \cite{Schneider}.
Here we sketch how these properties are taken into account,
{\it i.e.}, how to include
the light scattering of the optical black hole.
In our simple one-dimensional example, scattering occurs only
when the permittivity $\varepsilon$ of the medium varies
in space, because the solution (\ref{eq:axt}) for constant
$\varepsilon$ maintains co- and counter-propagating light
as the separate waves $A_+(\tau_+)$ and $A_-(\tau_-)$.
When $\varepsilon$ varies, we represent the monochromatic
modes on the two sides of the horizon as
\begin{eqnarray}
A_R & = & A_{+R} \exp({-i\omega\tau_+}) +
A_{-k} \exp({-i\omega\tau_-}) \,,
\nonumber\\
A_L & = & A_{+L} \exp({+i\omega\tau_+}) +
A_{-k} \exp({-i\omega\tau_-}) \,,
\label{eq:astructure}
\end{eqnarray}
where the $A_{+R}$,  $A_{+L}$ and $A_{-k}$ are slowly varying
functions of $x$.
Similar to Sec.\ 2.3, we construct
the transfer matrix (\ref{eq:connex}) in order to connect
$A_{+R}$ and $A_{-k}$ between the various points
on the right side of the horizon.
Also in the case of moving media,
the transfer matrix exhibits the structure (\ref{eq:trans})
with the determinant (\ref{eq:det}).
The latter property is a consequence
of the conservation law (\ref{eq:consmove}) that implies
for monochromatic modes
\begin{equation}
\partial_x \Big(i(\varepsilon u^2 - c^2)
(A^*\,\partial_x A - A\,\partial_x A^*) +
2\omega\varepsilon u \,|A|^2 \Big) = 0 \,.
\end{equation}
Substituting the structure (\ref{eq:astructure})
with the transfer-matrix connection (\ref{eq:connex})
gives the differential equation for the determinant (\ref{eq:det}).

Similar to Sec.\ 2.4, we use the transfer matrix to define
the in and out modes, but starting here with the outgoing modes,
\begin{equation}
A_R^\mathrm{out} = u_R(x)\,e^{-i\omega t} \,,\quad
A_D^\mathrm{out} = u_D(x)\,e^{-i\omega t} \,,
\end{equation}
where $A_R^\mathrm{out}$ describes an outgoing
counter-propagating wave,
whereas $A_D^\mathrm{out}$ refers to a co-propagating wave.
The $A_R^\mathrm{out}$ wave originates at the horizon,
propagating to the right,
whereas $A_D^\mathrm{out}$ propagates to the left at $+\infty$
and crosses the horizon without anything unusual happening.
We assume that at $+\infty$ the effective speed of light, $c'$ and
the flow $u$ approach constants such that light
approaches plane waves with the wavenumbers
\begin{equation}
k_R = \frac{\omega}{c'(+\infty) + u(+\infty)} \,,\quad
k_D = \frac{\omega}{c'(+\infty) - u(+\infty)} \,.
\end{equation}
In this case $u_R$ obeys the asymptotics
\begin{equation}
u_R(x) \sim {\cal A}_R
\left\{
\begin{array}{r@{\quad:\quad}l}
(x/x_{+\infty})^{i\omega/\alpha} & x \rightarrow +0\\
a^*\,e^{ik_Rx} + b^*\,e^{-ik_Dx}
& x \rightarrow +\infty
\end{array}
\right.
\end{equation}
and $u_D$ satisfies the corresponding asymptotics
similar to the complex conjugate of Eq.\ (\ref{eq:as1}).
On the other side of the horizon, we define an outgoing
counter-propagating wave by
\begin{equation}
A_L = u_L(x)\,e^{+i\omega t} \,,\quad
u_L(x) \sim {\cal A}_L \,(x/x_{-\infty})^{i\omega/\alpha}
\, \,:\, \, x \rightarrow -0 \,.
\end{equation}
We find that  $A_D^\mathrm{out}$
is orthogonal to  $A_R^\mathrm{out}$ with respect to
the scalar product (\ref{eq:scalarmove}) on the right side of the
horizon. Furthermore, $A_D^\mathrm{out}$ is automatically
orthogonal to $A_L^\mathrm{out}$ on the left side,
because here the two waves oscillate at opposite frequencies.
Consequently, the $A_R^\mathrm{out}$, $A_D^\mathrm{out}$,
$A_L^\mathrm{out}$ establish a set suitable for the
mode expansion (\ref{eq:modeex}) of quantum light,
the set of the outgoing modes.
The ingoing modes are the two analytic vacuum modes
$A_1$ and $A_2$ of Eq.\  (\ref{eq:bogo}), and a mode
$A_D^\mathrm{in}$
that is connected to $A_R^\mathrm{out}$ and $A_D^\mathrm{out}$
by the scattering matrix $\underline{B}$
of Eq.\ (\ref{eq:mediab}) on the right side of the horizon.
We thus arrive at the transformations
\begin{equation}
\left(
    \begin{array}{c}
     \hat{a}_R^\mathrm{out}  \\
     \hat{a}_D^\mathrm{out}
    \end{array}
\right)
=
\underline{B}
\left(
    \begin{array}{c}
     \hat{a}_R^\mathrm{in}  \\
     \hat{a}_D^\mathrm{in}
    \end{array}
\right)\,.
\end{equation}
Consequently, the
performance of the black hole as a
passive optical instrument is entirely due to
the light scattering on the right side of the horizon.
The scattering matrix describes the fraction
(\ref{eq:mediab}) of the generated
Hawking radiation that reaches the detector at $x=+\infty$.
The rest of the generated light is scattered into
co-propagating waves across the horizon and lost.
The black-body radiation of the hole is thus moderated
by the grey-body factor $a^{-1}\sqrt{|a|^2-|b|^2}$.

\section{Summary}

We have gone a long way from studying a piece of glass to the quantum
physics of black holes. Objects as diverse as beam splitters,
multiports, interferometers, fibre couplers, polarizers, gravitational lenses,
parametric amplifiers, phase-conjugating mirrors and
black holes have something in common ---
they act on light as simple optical instruments, {\it i.e.}\
as linear optical networks.
A linear optical network turns a set of incident light
modes into an equal number of outgoing modes by a linear
transformation. If the transformation does not mix annihilation
and creation operators the instrument is passive and it does conserve
the total number of photons. Otherwise, the instrument is active
and requires an energy source (or drain).
Such optical networks are reversible devices where, in principle,
all the outgoing modes can be reversed to restore the incident modes.
However, we can also model irreversible devices such as
absorbers or amplifiers by fictitious beam splitters or
parametric amplifiers, respectively, where we keep track
of only one of the outgoing modes. The quantum noise of
the unobserved modes accounts for the absorption or amplification
noise and the mode itself represents the absorption or
amplification reservoir.

Our starting point has been the quantum optics of a dielectric slab,
say a piece of glass. This simple example indicates
how to develop a quantum theory of light in dielectric media
and it shows how the mode transformations of passive
optical instruments come about.
We have developed the theoretical tools to analyze the quantum physics
of linear optical networks and we have applied them to a few
characteristic situations, to the splitting and interference of photons
and to the manifestation of quantum correlations in
parametric downconversion.
We have sketched how to describe irreversible processes
in quantum mechanics and how to apply this theory to determine
effective models for absorbers and amplifiers.
Finally, we returned to the starting point, revisiting the quantum
optics in dielectrics. We showed that the creation of a horizon
turns a passive medium into an active device, similar to
the quantum black hole.

\section*{Acknowledgments}

I thank
John Allen,
Enrique Arilla,
Conor Farrell,
Igor Jex,
Natasha Korolkova,
Paul Kwiat,
Irina Leonhardt,
Arnold Neumaier,
Renaud Parentani,
Harry Paul,
Stig Stenholm,
Ilya Vadeiko,
Gregor Weihs
and
Anton Zeilinger
for helping me with this article.
Many other people have contributed to my
understanding of the quantum physics of simple
optical instruments.
I am grateful for the financial support of the
Leverhulme Trust,
the ESF Programme Cosmology in the Laboratory
and the
Engineering and Physical Sciences Research Council.



\end{document}